\documentclass[notitlepage, final, 10pt, pra, aps, twocolumn, floatfix, showpacs, letter]{revtex4-1}
\pdfoutput=1

\usepackage{lmodern}
\usepackage[english]{babel} 
\usepackage{graphicx}
\usepackage{color}  

\definecolor{navyblue}{rgb}{0.0, 0.0, 0.5} 
\usepackage{amsmath, amssymb, amsfonts,bm}
\usepackage{latexsym}   
\usepackage{bbm} 
\usepackage{mathtools}  
\usepackage{placeins}   
\usepackage{braket}  
\usepackage[protrusion=true, expansion=true]{microtype}
\usepackage[colorlinks=true, breaklinks=false, linkcolor=navyblue, urlcolor=navyblue, citecolor=navyblue]{hyperref}

\newcommand{\ot}[1]{\hat{#1}} 
 
\newcommand{\dd}[1]{\text{d}#1}
\newcommand{\D}[1]{\delta #1} 
\newcommand{\ii}[0]{\mathrm{i}} 
\newcommand{\ee}[0]{\mathrm{e}}
\newcommand{\sigmap}[0]{{\sigma'}}
\newcommand{\ua}[0]{\uparrow} 
\newcommand{\da}[0]{\downarrow}
\newcommand{\updownarrows}{\mathbin\uparrow,\hspace{-.0em}\downarrow}
\DeclareMathOperator{\Tr}{Tr}

\begin{document}
\title{New class of quantum error-correcting codes for a bosonic mode}
\author{Marios H.~Michael}\altaffiliation{Present address: Cavendish Laboratory, University of Cambridge, JJ Thomson Avenue, Cambridge CB3 0HE, UK}
\author{Matti Silveri}\altaffiliation{Present address: Theoretical Physics, University of Oulu, Finland} \email{matti.silveri@oulu.fi} 
\author{R.~T.~Brierley} 
\author{Victor~V.~Albert} 
\author{Juha Salmilehto} 
\author{Liang Jiang} 
\author{S.~M.~Girvin}
\affiliation{Departments of Physics and Applied Physics, Yale University, New Haven, Connecticut 06520, USA}
\pacs{03.67.Pp, 
42.50.Ex, 
85.25.Cp, 
}
\date{\today}
\selectlanguage{english}

\begin{abstract}We construct a new class of quantum error-correcting codes for a bosonic mode which are advantageous for applications in quantum memories, communication, and scalable computation. These `binomial quantum codes' are formed from a finite superposition of Fock states weighted with binomial coefficients. The binomial codes can exactly correct errors that are polynomial up to a specific degree in bosonic creation and annihilation operators, including amplitude damping and displacement noise as well as boson addition and dephasing errors. For realistic continuous-time dissipative evolution, the codes can perform approximate quantum error correction to any given order in the timestep between error detection measurements. We present an explicit approximate quantum error recovery operation based on projective measurements and unitary operations. The binomial codes are tailored for detecting boson loss and gain errors by means of measurements of the generalized number parity. We discuss optimization of the binomial codes and demonstrate that by relaxing the parity structure, codes with even lower unrecoverable error rates can be achieved. The binomial codes are related to existing two-mode bosonic codes but offer the advantage of requiring only a single bosonic mode to correct amplitude damping as well as the ability to correct other errors. Our codes are similar in spirit to `cat codes'  based on superpositions of the coherent states, but offer several advantages such as smaller mean boson number, exact rather than approximate orthonormality of the code words, and an explicit unitary operation for repumping energy into the bosonic mode. The binomial quantum codes are realizable with current superconducting circuit technology and they should prove useful in other quantum technologies, including bosonic quantum memories, photonic quantum communication, and optical-to-microwave up- and down-conversion.
\end{abstract}

\maketitle

Continuous variable quantum information processing using bosonic modes~\cite{Braunstein98, Lloyd98, Braunstein05, Niset08, Aoki09, Lassen10, Weedbrook12_review, Menicucci14} offers an attractive alternative to two-level qubits. In practice, hybrid systems containing fixed qubits, photons trapped in resonators, as well as flying photon qubits will likely be essential to realistic architectures for quantum computation and communication~\cite{WiringUp08,Devoret13,Brecht15}. There is lively current interest in novel schemes for robustly encoding quantum information in bosonic modes~\cite{Chuang97, Gottesman01, Glancy04, Wasilewski07, Cochrane99, Leghtas13, Mirrahimi14, albert_holonomic_15, Terhal15, Grassl_2015, Bergmann15}. In particular, early work by Chuang~\textit{et~al.}~\cite{Chuang97} developed two-mode bosonic codes which can protect quantum information against amplitude damping of the bosonic field. This paper presents a new class of codes that are similar but have two distinct advantages: they require only a single bosonic mode (not two with identical decay rates), and they can correct other errors, \textit{e.g.},\  dephasing in addition to amplitude damping. Our codes are also similar in spirit to the `cat codes'~\cite{Cochrane99, Leghtas13, Mirrahimi14, albert_holonomic_15} and indeed asymptotically approach them in certain parameter limits.  However our codes require a smaller mean boson number to achieve a given fidelity, and are defined in a finite Hilbert space, making explicit construction of the required unitary operations more straightforward.

Besides extending the lifetimes of bosonic qubits or quantum memories, we describe how such codes could be useful in increasing the fidelity of quantum communication and remote entanglement between hardware modules through photon `pitch and catch' protocols~\cite{Zoller97, Wenner14, Houck14} as well as in improving the fidelity of communication based on reversible microwave to optical state conversion~\cite{Konrad-Cindy14}. In general, a bosonic mode can refer to an electromagnetic, magnetic or mechanical mode. However, the most likely implementations in the short term are photonic, so throughout this paper we will use term photon than the more general boson, for simplicity. These codes are of general interest for quantum information processing but we will place particular emphasis on the possibility of realizing them and the Chuang~\textit{et~al.}\ code~\cite{Chuang97} using superconducting qubits dispersively coupled to microwave resonators. Given the remarkable experimental progress in circuit QED in the last decade~\cite{Devoret13, WiringUp08}, such codes are ideally suited for this architecture and should be feasible with current technology. 

Superconducting qubit phase coherence times have risen some five orders of magnitude and can now reach $\sim 100\ \mu\mathrm{s}$ in 3D cavity geometries~\cite{Paik2011,IBM3D} and up to $\sim 40\ \mu\mathrm{s}$ in planar geometries~\cite{ChangPappas13,XMONPhysRevLett.111.080502,YanOliver15}. Superconducting microwave resonators are simpler than Josephson junction based qubits and can readily be constructed to have lifetimes even longer than the best qubits~\cite{Bertet02, Haroche65ms, Devoret13, Reagor10msCavity, ReagorMillisecondMemory}. It is therefore interesting to consider the possibility of using microwave resonators as qubits or as quantum memories. Experimental capabilities to arbitrarily control the quantum state of a hybrid cavity-qubit system are now so advanced~\cite{Houck_single_photon07, Hofheinz09, 100photonCat, Sun14,  Bell_the_Cat, Krastanov15, Heeres15, ChenWang16, Nissim16} that it is time to take the next step and  develop optimal quantum error correction codes that further extend the lifetimes of photonic quantum bits and memories. 
 
Because a resonator mode is a simple harmonic oscillator with equidistant level spacing, the only quantum states that can be accessed via classical linear drives are simple coherent states. To create more useful quantum superpositions of photon Fock states which can store quantum information, it is necessary to couple the bosonic mode to a non-linear element, \textit{e.g.},\ a superconducting qubit~\cite{Heeres15, Krastanov15}, a trapped ion~\cite{Chiaverini04, WinelandScience, Schindler11, Nigg14} or a Rydberg atom~\cite{Bertet02,Haroche65ms}. Experiments have demonstrated coherent mapping of a superconducting qubit state onto the corresponding coherent superposition of $0$ and $1$ photons in a resonator~\cite{Houck_single_photon07} and the creation of more complex superpositions~\cite{Hofheinz09}. Later experiments with superconducting qubits and cavities demonstrated the ability to transfer the qubit state into the cavity for a period $\sim 750~\mu$s much longer than the qubit lifetime and then return it to the qubit prior to verification via process tomography~\cite{ReagorMillisecondMemory}. Remarkably, it is also now possible to make quantum non-demolition measurements of photon number parity~\cite{Sun14} which can be used to greatly simplify the measurement of Wigner functions~\cite{Bertet02,100photonCat,Sun14}.  Such parity measurements are utilized as the error syndrome measurement in cat codes~\cite{Sun14,Mirrahimi14,Leghtas13} and will be used in some of the codes presented here. More generally, by employing optimal control pulses~\cite{Khaneja05, Fouquieres11} for engineering the driving terms of both a harmonic oscillator and a coupled non-linear element, one can realize arbitrary unitary operations for the full system with high experimental fidelity~\cite{Heeres15, Krastanov15, ChenWang16, Nissim16}. The practicality of the codes presented here relies on recent experimental progress in realizing this arbitrary unitary control.

Rudimentary quantum error correction~(QEC) protocols have been successfully carried out in ion traps~\cite{Chiaverini04,Schindler11,Nigg14}, with nuclear magnetic resonance~\cite{Cory98, Moussa11, ZhangLaflamme12}, with nitrogen-vacancy centers in diamond~\cite{Wadlherr14, Taminiau14}, Rydberg atoms~\cite{Haroche65ms} and superconducting qubits~\cite{Reed12,Kelly15, Corcoles15, Riste15}. However, it is difficult to reach the `break-even point', where the coherence time of the logical qubit exceeds the lifetime of the constituent physical qubits. This has recently been achieved in cavity photonic logical qubits~\cite{Nissim16} against photon loss errors and in nitrogen-vacancy centers in diamond~\cite{Cramer15} against dephasing errors. For qubit-based technologies, reaching break-even is challenging because it takes $N = 5$ physical qubits for the smallest possible code~\cite{Laflamme96, Bennett96, Gottesman09, Chuang_and_Nielsen}, $N = 7$ qubits for the Steane code~\cite{Steane96}, and $N = 9$ qubits for the Shor code~\cite{Shor95}. If the errors are uncorrelated single-qubit errors, the bare error rate is $N$ times faster than for a single qubit. Thus the quantum error correction protocol must overcome this factor of $N$ in order to reach break-even. If the errors are highly correlated, \textit{e.g.}\ in the case of uniform magnetic field fluctuations in an ion trap, there may exist a decoherence-free-subspace encoding which will be advantageous~\cite{ZanardiDFS,LidarDFS, BoulantDFS, Chuang_and_Nielsen, Schindler11}. To date there is no evidence that correlated errors are a significant problem for well-shielded superconducting qubits.

Another potential advantage of resonators is that the error model appears to be very simple. The dominant source of decoherence is the energy loss of the cavity as it slowly emits photons into its output and input ports. The cavity energy ring-down rate $\kappa$ is the analog of $1/T_1$ for a qubit. To date, there is no evidence of dephasing errors associated with frequency fluctuations of 3D metallic superconducting cavities though dephasing has been seen in coplanar waveguide resonators by the Zmuidzinas group~\cite{Zmuidzinas} and attributed to two-level systems in the dielectric substrate. However dispersive coupling of a qubit to a resonator will introduce random fluctuations of the cavity frequency associated with $T_1$ state change events in the qubit~\cite{ReagorMillisecondMemory}. Additionally, there can be, for example, energy leakage from the driven ancillary qubit to the resonator that can be result in photon gain errors. We will focus on correcting cavity photon loss errors but the constructed codes can be protected also against dephasing and photon gain errors as well. 

The paper is organized as follows. In Sec~\ref{sec:qec}, we introduce quantum error correction against discrete photon loss, photon gain and dephasing errors through simple bosonic single-mode codes. These are generalized to binomial quantum codes in Sec.~\ref{sec:binomcodes}. Next we consider realistic continuous-time dissipative evolution under these errors. Since the continuous-time evolution introduces an infinite set of errors even during a finite timestep, exact quantum error correction is impossible. In Sec.~\ref{sec:aqec}, we prove our main result\----the binomial quantum codes can perform approximate quantum error correction to any given order in timestep for realistic continuous-time dissipative evolution. We introduce an explicit and experimentally relevant recovery process. In the remainder of the paper, we analyze the performance of the codes, present comparisons to related pre-existing codes, and discuss applications in quantum communication and as logical qubits, respectively in Secs.~\ref{sec:perfor}-\ref{sec:logical}. Further improvements and overall discussion of the binomial codes are presented in Sec.~\ref{sec:discuss} before the summary and conclusion of Sec.~\ref{sec:conc}.

\section{Quantum error correction against photon loss, gain and dephasing errors} \label{sec:qec}
The generic task of quantum error correction is to find two logical code words\----a qubit\----embedded in a large Hilbert space. The code words are required to be robust such that if any one of the single, independent errors $\ot E_k \in \mathcal{\bar{E}}$ occurs, no quantum information is lost and any quantum superposition of the logical code words can be faithfully recovered. This is equivalent to finding two logical code words $\ket{W_\sigma}$, where $\sigma=\updownarrows$, that satisfy the quantum error correction criteria~\cite{Chuang_and_Nielsen, Terhal15_review}, known also as the Knill-Laflamme conditions~\cite{Bennett96, Knill97}, 
\begin{equation}
  \label{eq:qecc}
\braket{W_\sigma|\ot E_\ell^\dagger \ot E_{k}|W_\sigmap}=\alpha_{\ell k}\delta_{\sigma \sigmap},
\end{equation}
for all $\ot E_{\ell,k}\in \mathcal{\bar{E}}$ such that $\alpha_{\ell k}$ are entries of a Hermitian matrix and independent of the logical words. The independence of entries $\alpha_{\ell k}$ from the logical code words and the structure of the non-diagonal entries guarantee that the different errors are distinguishable and correctable.

We consider a damped harmonic oscillator suffering from photon losses. Our intention is to design logical code words, which are embedded in the Hilbert space of the harmonic oscillator and protected up to $L$ photon loss events occurring in the time interval $\D{t}$ between two consecutive quantum correction stages. The set of discrete errors is $\bar{\mathcal{E}}_L=\big \{\ot I, \ot a, \ot a^2, \ldots \ot a^L \big \}$~\footnote{Notice that this is not a completely positive and trace preserving (CPTP) map, completion to such will be done in Sec.~\ref{sec:aqec}.}. We will see that satisfying the conditions \eqref{eq:qecc} for these errors is sufficient to produce a code that is systematically correctable up to a given order in $\kappa \D{t}$ under the full amplitude damping operators. Here our primary focus will be on photon loss errors. Later we will discuss correcting photon gain $\ot a^\dagger$ and dephasing errors $\ot n$.

Suppose we restrict our attention to the lowest $2^M$ Fock states of a single mode of a resonator. This Hilbert space is the same size as that for $M$ qubits and if one had complete control over it, one could imagine having a kind of hardware shortcut in which a single resonator mode replaces a complex of $M$ physical qubits to form one or more logical qubits~\cite{Devoret13}. Notice that increasing the size of the Hilbert space does not increase the number of error channels or the minimal number of error syndromes. Consider the following simple encoding of $M$ qubits into the state of the resonator. The $2^M$ Fock states cover photon numbers $0,1,\ldots, (2^M-1)$. Let photon Fock state $\ket{n}$  be represented by $\ket{n}=\ket{b_{M-1}^{n}b_{M-2}^{n}\ldots b_1^{n} b_0^{n}}$, where $b_{M-1}^{n}b_{M-2}^{n}\ldots b_1^{n} b_0^{n}$ is the binary representation of the number $n$. The $j$th binary digit represents the eigenvalue  $(1+\hat{\sigma}_j^z)/2$ for the corresponding `physical qubit.' This appears to be a very simple and satisfactory encoding but consider what happens when the $n = 8$ state (say) loses a single photon $\ot a \ket{1000}=\sqrt{8} \ket{0111}$. What seems to be a simple error model in terms of photon loss actually becomes a model with correlated multi-qubit errors. Hence typical quantum error correction schemes based on models of independent single qubit errors cannot be easily transferred to this problem~\cite{Gottesman09, Chuang_and_Nielsen}. Because the mean photon loss rate $\kappa\langle \ot a^\dagger \ot a\rangle$ scales exponentially with $M$, we will focus here on representing a single logical qubit using a small number of states in the cavity to permit error correction. 

An example of a code protecting against $\bar{\mathcal{E}}_1=\big\{\ot I, \ot a \big\}$ is
  \begin{align}
\ket{W_\ua}&=\frac{\ket{0}+\ket{4}}{\sqrt{2}}, & \ket{W_\da}&=\ket{2}. \label{eq:042}
\end{align}
A photon loss error brings the logical code words to a subspace with odd photon numbers that is clearly disjoint from the even-parity subspace of the logical code words. Therefore, the off-diagonal parts of the quantum error correction matrix~\eqref{eq:qecc} $\alpha$ are identically zero. The remaining diagonal part of $\alpha$ requires that the mean photon number is identical for both of the states, here $\bar{n}=\braket{W_\sigma|\ot n|W_\sigma}=2$. This means that the probability of a photon jump to occur (or not to occur) is the same for both of the states, implying that the quantum state is not deformed under an error. Explicitly, if a quantum state $\ket{\psi}=u \ket{W_\ua}+ v \ket{W_\da}$ suffers a photon jump, it is transformed to $\ket{\psi_1}=\ot a \ket{\psi}/\sqrt{\Braket{\psi|\ot a^\dagger \ot a|\psi}}= u \ket{\bar{E}^1_\ua}+v \ket{\bar{E}^1_\da}$, where $\ket{\bar{E}^1_\ua}=\ket{3}$ and $\ket{\bar{E}^1_\da}=\ket{1}$ denote the error words. The quantum information (the complex coefficients $ u $ and $ v $) is not deformed. 

In the optical regime, the photon loss error can be detected by an external photodetector.  This is not yet practical in the microwave regime, but fortunately we have the capability of very high fidelity quantum non-demolition measurements of the photon number parity~\cite{Sun14}. The original state is recovered by a unitary operation $\ot U_1$ that performs the state transfer $\ket{\bar{E}^1_\sigma} \leftrightarrow \ket{W_\sigma}$; see Appendix~\ref{app:binom_unitary} for details. Correcting a non-unitary error with a unitary operation works only because it is conditioned on the detection outcome of the particular error. Notice that the code~\eqref{eq:042} is similar to a code developed in Ref.~\cite{Chuang97} by Chuang \textit{et al.}\ for a multi-mode system. Our code has the important advantages of requiring only a single bosonic mode and having rate for uncorrectable errors that is smaller by a factor of three (see Sec.~\ref{sec:comparison}).

Generalizations of the code~\eqref{eq:042} are possible, for example \begin{align}
\ket{W_\ua}&=\frac{\ket{0}+\sqrt{3}\ket{6}}{2}, & \ket{W_\da}&=\frac{\sqrt{3}\ket{3}+\ket{9}}{2}. \label{eq:0639}
\end{align}
In addition to the logical words having the same mean photon number, the error words $\ket{\bar{E}^1_\sigma}=\ot a \ket{W_\sigma}/\sqrt{\braket{W_\sigma|\ot a^\dagger \ot a|W_\sigma}}$, $\ket{\bar{E}_\ua^1}=\ket{5}$ and $\ket{\bar{E}_\da^1}=(\ket{2}+\ket{8})/\sqrt{2}$ also have the same mean photon number. Thus the code can tolerate another photon loss error and the protected error set is $\mathcal{\bar{E}}_2=\big \{ \ot I, \ot a, \ot a^2 \big\}$. The photon loss errors are detected by measuring photon number $\bmod\ 3$; see Appendix~\ref{app:binom_unitary}. The error recovery procedure is similar to that above: an error detection is followed by a unitary operation performing a state transfer $\ket{W_\sigma} \leftrightarrow \ket{\bar{E}^k_\sigma}$.
The error correction matrix $\alpha$ for the code~\eqref{eq:0639} is diagonal because of the sufficiently large spacing of the occupied Fock states in the logical words such that photon loss errors $\ot a^\ell$, $\ell\le 2$, cannot lead to overlap of the error words.

Frequency fluctuations of the cavity (\emph{i.e.}\ noise $\xi(t)$ coupling to the photon number, $\xi(t)\ot n$, for example by transitions of a dispersively coupled ancilla qubit~\cite{ReagorMillisecondMemory}) cause dephasing of the quantum memory. In the limit of fast Markovian noise, the effect of fluctuations is well approximated by a Lindblad dissipator with a jump operator $\sqrt{\gamma}\ot n$, where $\gamma$ is the dephasing rate. Even in the case of non-Markovian noise, the errors take the form $\ot U(\D{t})=\exp(-\ii \int_0^{\D{t}}\xi(\tau)d{\tau} \ot n)$. For small enough timesteps $\D{t}$ this error can be expanded as superposition of $\ot n^k$ operators. Thus in both cases protection can be achieved by considering the operator $\ot n$ and its higher powers. In what follows we will refer to these operators as dephasing errors.  

Due to the spacing of the Fock states and the properties of bosonic operators, the code~\eqref{eq:0639} protects also against a dephasing error $\ot n$, thus the full error set is $\mathcal{\bar{E}}_2=\big \{ \ot I, \ot a, \ot a^2, \ot n \big \}$. Since the dephasing error does not change the photon number, it leads to an error state $\ket{\psi_n}=\ot n \ket{\psi}/\sqrt{\braket{\psi|\ot n^2|\psi}}$,
\begin{align}
\ket{\psi_n} =u\frac{\sqrt{3}\ket{W_\ua}-\ket{\bar{E}^{n}_\ua}}{2}+v\frac{\sqrt{3}\ket{W_\da}-\ket{\bar{E}^{n}_\da}}{2}. \label{eq:n-error}
\end{align}
which is a superposition of the original words and the error words related to the dephasing $\ket{\bar{E}^{n}_\ua}=(\sqrt{3}\ket{0}-\ket{6})/2$ and $\ket{\bar{E}^{n}_\da}=(\ket{3}-\sqrt{3}\ket{9})/2$. The only way to detect the dephasing error is to make projective measurements into the logical word basis $\ot P_{\rm W}=\sum_\sigma \ket{W_\sigma}\bra{W_\sigma}$ and if the answer is negative and no photon loss errors were detected, the original state is recovered by making a unitary operation performing a state transfer $\ket{\bar{E}_\sigma^n}\leftrightarrow \ket{W_\sigma}$. Remarkably such complex operations applied to a cavity-ancilla qubit systems are now technically feasible~\cite{Heeres15, Krastanov15, ChenWang16, Nissim16}.

The code~\eqref{eq:0639} can instead be chosen to be protected against errors $\mathcal{\bar{E}}'_2=\big \{ \ot I, \ot a, \ot a^\dagger, \ot n \big \}$ since a photon gain error and two-photon loss errors have the same change in the photon number $\bmod\ 3$ and the logical code words already obey the quantum error correction condition for the photon gain error: $\braket{W_\sigma|\ot a \ot a^\dagger|W_\sigmap}=(\bar{n}+1)\delta_{\sigma \sigmap}$. As a special case, one can choose to protect only against $\mathcal{\bar{E}}'_1=\big \{\ot I, \ot a, \ot n \big\}$ achieved by the same Fock state coefficients as with the code~\eqref{eq:0639} but with spacing of the code~\eqref{eq:042}: 
\begin{align}
\ket{W_\ua}&=\frac{\ket{0}+\sqrt{3}\ket{4}}{2}, & \ket{W_\da}&=\frac{\sqrt{3}\ket{2}+\ket{6}}{2}. \label{eq:0426}
\end{align}
The relationship between the codes~\eqref{eq:0639} and \eqref{eq:0426} arises from a general structure, which we exploit in the next section.

\section{Binomial quantum codes}\label{sec:binomcodes}
\begin{figure}
\centering
\includegraphics[width=1.0\linewidth]{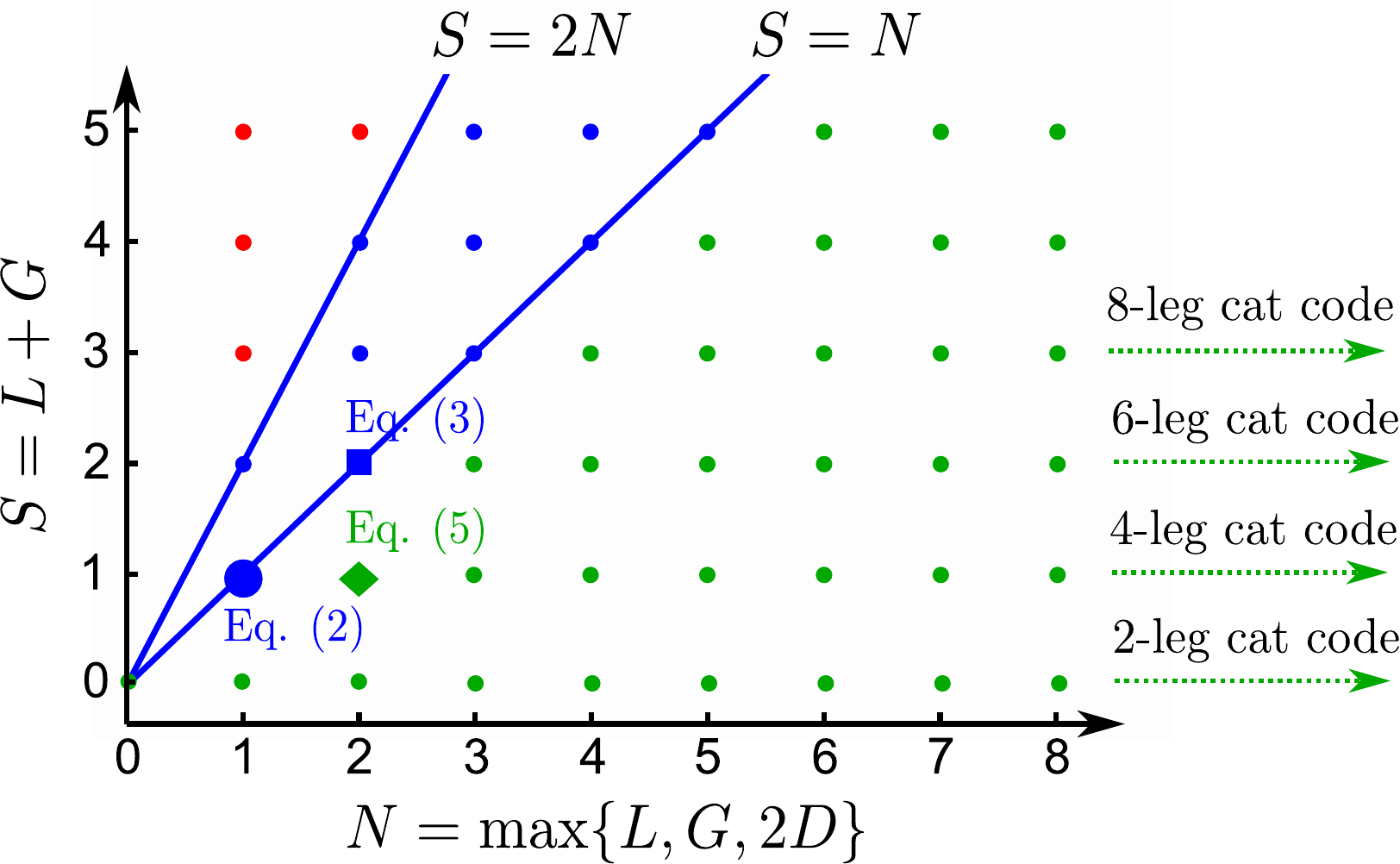}
\caption{\label{LDdfigure}Two-parameter~$(N,S)$ space of the binomial codes~\eqref{eq:binomial}. The largest blue circle denotes the code~\eqref{eq:042} protected against a photon loss error $L=1$, the blue square is the code~\eqref{eq:0639} protected against $\mathcal{\bar{E}}_2=\{\ot I, \ot a, \ot a^2, \ot n  \}$ or $\mathcal{\bar{E}}'_2=\{\ot I, \ot a, \ot a^\dagger, \ot n \}$, and the green diamond denotes the code~\eqref{eq:0426} protected against $\mathcal{\bar{E}}'_1=\{\ot I, \ot a, \ot n \}$. The parameter $S=L+G$ sets the total number of detectable photon loss $L$ and gain $G$ errors. The parameter $N$ sets the maximum order the code is protected against photon loss, gain and dephasing errors  $N=\max\left\{L, G, 2D \right\}$. The codes denoted with blue have protection against photon loss and gain errors set by $S=L+G$ and in addition they are protected against dephasing up to $\ot n^{\lfloor \max \{L, G\}/2 \rfloor}$. The codes denote with red allow in addition heralding of $S-2N$ uncorrectable photon loss or gain errors. The codes denoted with green are protected against a total of $S$ photon loss and gain errors, as well as against up to $\ot n^{N/2}$ dephasing errors. In the limit $N\to \infty$ the binomial codes asymptotically approach the $2(L+1)$-legged cat codes~\cite{Leghtas13, Mirrahimi14, albert_holonomic_15} protected against $L$ photon loss errors.}
\end{figure}

We now  generalize the above codes to protect against the error set,
\begin{equation}
\mathcal{\bar{E}}=\big\{ \ot I, \ot a, \ot a^2, \ldots, \ot a^L, \ot a^\dagger, \ldots, (\ot a^\dagger)^G,  \ot n, \ot n^2, \ldots, \ot n^D  \big \}, \label{eq:lll}
\end{equation}
which includes up to $L$~photon losses, up to $G$~photon gain errors, and up to $D$~dephasing events. We have found a simple class of codes which can correct such an error set, 
\begin{equation}
  \ket{W_{\ua/\da}}=\frac{1}{\sqrt{2^{N}}} \sum_{\text{p even/odd}}^{[0, N+1]} \sqrt{\binom{N+1}{p}}\ket{p(S+1)}, \label{eq:binomial}
\end{equation}
where the spacing is $S=L+G$, maximum order $N=\max\left\{L, G, 2D\right\}$ and the range of the index $p$ is from $0$ to $N+1$. The two-parameter $(N, S)$ code space is shown in Fig.~\ref{LDdfigure}. Because the Fock state coefficients involve binomial coefficients we refer to this class as \emph{the~binomial~codes}.

The spacing between the occupied Fock states is $S+1$ such that the correctable $L$ photon loss and $G$ gain errors can be uniquely distinguished by measuring photon number modulo $S+1$, which we call 'generalized parity' here. The  quantum error correction conditions~\eqref{eq:qecc} require that $\braket{W_\sigma|(\ot a^\dagger)^\ell \ot a^\ell |W_\sigma}$, for all $\ell\le \max\{L, G\}$, is equal for the two logical code words. Satisfaction of Eq.~\eqref{eq:qecc} guarantees that the quantum state is not deformed under an error and implies as well an existence of a recovery process\----the detectable errors can be recovered using unitary operations. By using commutation relations it is equivalent to require that $\braket{W_\sigma|\ot n^\ell |W_\sigma}$, for all $\ell\le \max\{L, G\}$, be equal for the two logical code words, just as the mean photon number of the logical code words~\eqref{eq:042} was required to be equal.  A straightforward way of seeing that expectation values for moments of the photon number are equal is to recall the binomial formula and consider the difference 
\begin{align}
\Delta_\ell&=\braket{W_\ua|\ot n^\ell |W_\ua}-\braket{W_\da|\ot n^\ell |W_\da}\notag\\
&=\frac{(S+1)^\ell}{2^N}\sum_{p=0}^{N+1}\binom{N+1}{p} p^\ell (-1)^p\notag \\
&=\frac{(S+1)^\ell}{2^N}\left. \left(x\frac{\dd{}}{\dd x}\right)^{\ell}(1+x)^{N+1}\right|_{x=-1}\,. \label{eq:ndiff}
\end{align}
The derivative on the last line preserves at least one $(1+x)$ in each of the terms of the above polynomial such that $\Delta_{\ell}=0$ for all $\ell\le N=\max\{L, G\}$; see Appendix~\ref{app:binom}. It should be noted that the coefficients of the Fock states in the code words~\eqref{eq:binomial} are independent of the spacing $S+1$. The spacing between occupied Fock states enables the detection of photon loss errors. The values of the Fock state coefficients are determined by balancing the moments of the photon number distribution so that the rate of errors is the same for all logical states.

The basis of the two logical code words can be generalized to $d$ logical code words, a so-called `qudit', by utilizing extended binomial coefficients $\binom{N+1}{p}_d $~\cite{Neuschel2014, extended_binomial} (defined in Appendix~\ref{app:qudit}; these are also called polynomial coefficients \cite{Caiado2007}),
\begin{align}
  \ket{W_i}=\frac{1}{\sqrt{d^N}} \sum_{p = i \bmod d}^{[0, p_{\rm m}]} \sqrt{\binom{N+1}{p}_d}\ket{p(S+1)}, 
\end{align}
where $i=0,1,\ldots, d-1$ and $p_{\rm m}=(d-1)(N+1)$. By using a similar argument as with the binomial qubit codes in Eq.~\eqref{eq:ndiff}, one can show that the photon number moments are equal for all of the $d$ code words; see details in Appendix~\ref{app:qudit}. In what follows we, for simplicity, concentrate on the binomial qubit codes but all the results can be extended to the binomial qudit codes as well.  

For measuring photon number~$\bmod\ S+1$ the mode needs to be coupled to an ancillary system, either a superconducting qubit~\cite{Heeres15, Krastanov15}, a trapped ion~\cite{Chiaverini04, WinelandScience, Schindler11, Nigg14} or a Rydberg atom~\cite{Bertet02,Haroche65ms}.  With superconducting circuits, the parity could be measured by utilizing the strong dispersive  coupling between the cavity and the ancillary qubit and by using number selective coherent drives similar to the demonstrated quantum non-demolition parity measurement~\cite{100photonCat, Sun14, Heeres15, Krastanov15}; see Appendix~\ref{app:binom_unitary}. Encoding the different outcomes of a multi-valued measurement, such as $S\ge 2 $, in the higher excited states of the ancillary qubit would allow single-shot measurements. An alternative is to do $S$ sequential measurements with a two-level qubit.

Dephasing errors result in mutually non-orthogonal error states: for them the quantum error correction matrix is non-diagonal but Hermitian. For that reason, to detect and recover those errors one needs to make projective measurements in an orthonormalized basis, as with the code~\eqref{eq:0639}. After a detection of an error, the original state is recovered by a unitary operation performing a state transfer between the subspaces of the error and logical code words.

\subsection{Errors correctable by binomial codes} \label{bin:err:cor}
The binomial code coefficients were derived using the requirement to be able to correct the photon loss or gain errors. Considering only photon loss errors, the parameter $L+1$ can be interpreted as the distance of the binomial quantum codes since it is the minimum number of $\ot a$ operators needed for mixing the two code words. Inclusion of dephasing errors makes the quantum error correction matrix~\eqref{eq:qecc} non-diagonal, but it automatically follows from the binomial coefficients that dephasing errors up to order $\lfloor \max \{L, G\}/2 \rfloor$ are also corrected by these codes. The highest degree of dephasing protection $\ot n^D$ does not need to be limited to the value set by the photon loss/gain protection. By increasing the length $N$ of the binomial code words, $D$ can be increased without a limit. This gives the maximum order as $N=\max\{L, G, 2D \}$. Note also that since the binomial codes are protected against any one of the errors from the error set~\eqref{eq:lll}, they are also protected against any errors that are superpositions of these. Such errors include, for instance, displacement `unitary' errors 
\begin{equation}
  \label{eq:displacement}
  \ot D(\beta)=\exp\left(\beta \ot a^\dagger-\beta^\ast \ot a\right)
\end{equation}
for small unknown $\beta$. More precisely, a binomial code with given values of $N$, $L$ and $G$ satisfies:
    \begin{equation}
      \label{eq:bin-conditions}
\braket{W_\sigma|(\ot a^\dagger)^{n_+}\ot a^{n_-} |W_\sigmap}
      =\begin{cases}
        0,&\text{if } n_+\neq n_-\text, \\
        \alpha_{n_+}\delta_{\sigma\sigmap}&\text{if } n_+ = n_- \leq N\text,
      \end{cases}
    \end{equation}
where $|n_+-n_-|\leq G+L$ with $n_+, n_-\geq 0$ and the $\alpha_{n_+}$ are constants. In particular, if
the condition is satisfied for some values of $n_+$ and $n_-$, it
is satisfied for \emph{all smaller values}.
This property means, for example, that the code can correct all errors
of the form
\begin{equation}
    \label{eq:high-order-possible-errors}
      \ot A_i = \sum_{jk}\xi^{(i)}_{jk}(\ot a^\dagger)^{j}\ot a^{k},
    \end{equation}
if all nonzero $\xi^{(i)}_{jk}$ satisfy $0\leq j-k\leq G$, $0\leq k-j\leq L$ and $j+k\leq N$. 
When choosing a code, the minimum value of $N$ is determined by the term with the
largest total number of $\ot a$ and $\ot a^\dagger$ operators. The 
parameter $L$ ($G$) is given by the error term that causes the largest overall decrease
(increase) in excitation number. For example, to correct $\ot A=\ot
a^\dagger + \ot n^2$ requires $G \ge 1$ from the first term, $N \ge 2$ from the
second term and $L \ge 0$ as there are no number-decreasing terms.

\section{Approximate quantum error correction under continuous-time dissipative evolution} \label{sec:aqec}
Up to now we have assumed that the cavity is subject to a finite set of discrete errors. In fact, the cavity evolves continuously in time. For example, the standard Lindblad time evolution of a density matrix $\ot \rho$ of a cavity coupled to a zero-temperature bath with a cavity energy decay rate $\kappa$ (represented in the frame rotating at the cavity frequency) is 
\begin{equation}
 \dd{\ot \rho} = \kappa
\dd{t}\left (\ot a\ot \rho \ot a^\dagger -\frac{\ot a^\dagger \ot a}{2}\ot \rho - \ot \rho \frac{\ot a^\dagger \ot a}{2}\right ). \label{eq:lind_a}
\end{equation}
In a finite time interval $\D{t}$, continuous-time evolution results in an infinite set of possible errors. Exact quantum error correction of the full set of errors is not possible. However, the probabilities of the errors scale with powers of $\kappa\D{t}$ and we can choose to correct only the most important errors in $\kappa\D{t}$. Formally, we will exploit the notion and theory of approximate quantum error correction~\cite{Leung97, Gottesman2005, Ng10, beny_general_2010, beny_perturbative_2011, mandayam_towards_2012, Ouyang14_GNU, Ouyang15_GNU, Grassl_2015}. We expand each error operator in powers of $\kappa\D{t}$ and choose to correct up to a given highest order. It is then enough to satisfy the quantum error correction criteria~\eqref{eq:qecc} only approximately such that the original state can be recovered with an accuracy given by the same highest order in $\kappa \D{t}$.  

Initially, we consider only photon loss errors due to cavity damping and extend the discussion for photon gain and dephasing processes later. One can `unravel' the Lindblad equation~\eqref{eq:lind_a} for photon loss errors by considering the conditional quantum evolution of the system based on the measurement record of a photomultiplier that clicks whenever a photon leaks out of the cavity. In this quantum trajectory picture~\cite{OpenQuantumSystems}, one views the first term in Eq.~\eqref{eq:lind_a} representing the photon loss jump of the system when the detector clicks $\ot \rho\rightarrow \ot a\ot \rho \ot a^\dagger$. This is not normalized because it includes the fact that the click probability is proportional to $\Tr(\ot a \ot \rho \ot a^\dagger)=\bar{n}$. The last two terms inside the brackets represent time evolution of the system under the non-Hermitian Hamiltonian $\ot{V}/\hbar=-\ii\frac{\kappa}{2}\ot a^\dagger \ot a $ when no photons are detected. We have previously omitted this no-jump evolution for simplicity, but will now examine this crucial part of the physical error process.

Much like a Feynman path integral, we can express the evolution of the density matrix from time $0$ to $\D{t}$ in terms of a sum over all possible trajectories with photon loss jumps occurring at all possible times during the finite time interval $\D{t}$. We express this time evolution as a completely positive and trace preserving (CPTP) process $\mathcal{E}=\{ \ot E_0, \ot E_1, \ot E_2, \ldots \}$:
\begin{equation}
  \ot \rho(\D{t})=\mathcal{E}(\ot \rho(0)) \equiv \sum_{\ell=0}^\infty \ot E_\ell \ot \rho(0) \ot E_\ell^\dagger, \label{eq:rho:el}
\end{equation} 
where $\ot E_\ell$ are Kraus operators encapsulating the time evolution generated by exactly $\ell$ photon losses and the no-jump evolution.  Remarkably, by integrating over all the possible jump times of exactly $\ell$ photon jumps during the time interval $\D{t}$, we can derive an exact analytic expression for $\ot E_\ell$~\cite{Ueda89, Lee94, Chuang97}, 
\begin{align}
  \ot E_\ell=\sqrt{\frac{(1-\ee^{-\kappa \D{t}})^\ell}{\ell !}}\ee^{-\frac{\kappa \D{t}}{2}\ot n} \ot a^\ell\text{.}  
\label{eq:kraus}
\end{align}
See Appendix~\ref{app:kraus-deriv} for more details of the derivation. An important feature of the damped simple harmonic oscillator is that the precise timing of the the photon jumps plays no role. This can also be seen  in the interchangeability of the order of the operators in Eq.~\eqref{eq:kraus}: $\exp(- \kappa \D{t}\ot n) \ot a^\ell=\exp( \kappa \D{t} \ell)\ot a^\ell \exp(-\kappa \D{t} \ot n )$. Taken together, when correcting against photon loss errors up to order $(\kappa \D{t})^L$, the correct set of errors the codes should be protected against is (the Taylor series expansion of) $\mathcal{E}_L=\{\ot E_0, \ot E_1, \ldots, \ot E_L\}$~\footnote{This is a CPTP map to the desired accuracy of $(\kappa \D{t})^L$.}. This set includes contributions of both the jump and no-jump parts of the non-unitary time evolution. 

The measurement backaction of observing no photon jumps is non-trivial. It reduces the relative probability of the higher occupied Fock states with respect to the lower ones, formally expressed with the factor $\exp\left(- \frac 12 \ot n \kappa \D{t} \right)$ in the error operators~\eqref{eq:kraus}. Others have addressed this by constructing multimode codes~\cite{Mabuchi96, Chuang97, Plenio97, Grassl_2015}. These codes avoid no-jump evolution by combining two or more physical elements with identical decay rates and constructing the logical code words such that they are superpositions of states with the same combined total excitation number. They are essentially entangled versions of the codes~\eqref{eq:042}-\eqref{eq:0426}; see Sec.~\ref{sec:comparison}. 

One of our key results is that certain single-mode codes, such as the binomial codes~\eqref{eq:binomial}, can correct the no-jump evolution to the same accuracy as photon loss errors. We illustrate this first to accuracy $\mathcal O (\kappa \D{t})$ before giving a general description to an arbitrary accuracy in powers~of~$\kappa \D{t}$. 

\subsection{Approximate quantum error correction to first order in $\kappa \D{t}$} \label{sec:aqec:first}

Let us consider the code~\eqref{eq:042}. When no photon loss is detected, the quantum state $\ket{\psi}=u \ket{W_\ua}+v\ket{W_\da}$ transforms under the no-jump evolution given by $\ot E_0=\exp\left(-\frac 1 2 \kappa \D{t} \ot n \right)$. Code~\eqref{eq:042} is protected against a photon loss error occurring with a probability to first order in $\kappa \D{t}$: $P_1=\braket{\ot E^\dagger_1 \ot E_1}=\kappa \D{t} \bar{n}+\mathcal{O}[(\kappa \D{t})^2]$. Thus, it is reasonable to consider the no-jump evolution to the same accuracy. Taking into account $\ot E_0$ and the normalization in the denominator of $\ket{\psi_0}=\ot E_0 \ket{\psi}/\braket{\psi|\ot E^\dagger_0 \ot E_0|\psi}^{1/2}$ to first order in $\kappa \D{t}$, we obtain
\begin{align}
\ket{\psi_0}=&\Big [1+\frac{\kappa\D{t}}{2}(\bar{n}-\ot n) \Big] \ket{\psi}+\mathcal{O}[(\kappa \D{t})^2] \notag \\
=&u \left( \ket{W_\ua}+\kappa \D{t} \ket{E_\ua^0} \right) +v \ket{W_\da}+ \mathcal{O}[(\kappa \D{t})^2]. \label{eq:no_jump_state}
\end{align}
Here $\ket{E^0_\ua}=(\ket{0}-\ket{4})/\sqrt{2}$ is the error word associated with the no-jump evolution. Notice that the logical code word $\ket{W_\da}=\ket{2}$ is unaffected by the no-jump evolution as its excitation number is equal to the mean photon number. 

The no-jump error causes deterministic evolution inside the subspace $\{\ket{W_{\ua}}, \ket{E^0_\ua} \}$ which can be inverted to the desired accuracy by applying a unitary operator,
\begin{align}
  \label{eq:no-jump_dt}
  \ot U_0=&\sin\kappa \D{t} \left(\ket{W_\ua}\bra{E^{0}_\ua} -\ket{E^{0}_\ua}\bra{W_\ua}\right)\\ +&\cos \kappa \D{t} \left(\ket{W_\ua}\bra{W_\ua} +\ket{E^{0}_\ua}\bra{E^{0}_\ua}\right)+\ket{W_\da}\bra{W_\da} + \ot {U}_{\rm res}. \notag
\end{align}
Here, $\ot U_{\rm res}$ is an arbitrary unitary operator on the subspace complementary to the logical and error subspaces in order to complete $\ot U_0$ to a unitary operator in the entire Hilbert space. It can be taken to be the identity of the complementary subspace. By combining detection-correction of both the errors, the total recovery process is $\mathcal{R}=\{ \ot R_0, \ot R_1\}$. The Kraus operators are $\ot R_k =\ot U_k \ot \Pi_{k \bmod 2 }$, where $\ot \Pi_{k\bmod 2}$ is a projection into the photon number subspace $k \bmod 2$ and the correction unitary $\ot U_1$ is introduced in the text after Eq.~\eqref{eq:042}. The recovery processes results in the original state to first order in $\kappa \D{t}$ as desired: 
\begin{align}
\mathcal{R}\left(\mathcal{E}(\ot \rho)\right)&=\sum_{k=0}^1 \ot R_k \left( \sum_{\ell=0}^\infty \ot E_l \ot \rho \ot E^\dagger_\ell\right)\ot R_k^\dagger \notag \\ &=\sum_{k=0}^1\sum_{\ell=0}^1 \ot R_k \ot E_\ell \ot \rho \ot E^\dagger_\ell \ot R^\dagger_k+\mathcal{O}[(\kappa \D{t})^2]   \label{eq:recovery2}\\
&=\sum_{\ell=0}^1 \ot R_\ell \ot E_\ell \ot \rho \ot E^\dagger_\ell \ot R^\dagger_\ell+\mathcal{O}[(\kappa \D{t})^2]=\ot \rho +\mathcal{O}[(\kappa \D{t})^2].\notag
\end{align}
Here we have first ignored all the parts $\ot E_{\ell \ge 2}$ of the error process whose effect is $\mathcal{O}[(\kappa \D{t})^2]$. On the next line we use the knowledge that the initial state has even parity and the photon loss error process $\ot E_\ell$ shifts photon number by $\ell$, formally $\ot R_k \ot E_\ell \ot \Pi_{0 \bmod 2}=\ot U_k\ot \Pi_{k\bmod 2} \ot E_\ell \ot \Pi_{0 \bmod 2} =\delta_{k\ell} \ot R_\ell \ot E_\ell \ot \Pi_{0\bmod 2}$, for $k,\ell\in \{0,1 \}$. 

Alternatively the recovery can be done by a measurement projecting to the subspace of logical code words~\cite{Cirac96} such that $\mathcal{R}_{\rm m}=\{ \ot P_{\rm W}, \ot U_1 (\ot I-\ot P_{\rm W}) \}$. Here, $\ot P_{\rm W}=\sum_\sigma \ket{W_\sigma}\bra{W_\sigma}$ is the projection to logical subspace. If the measurement projects out of the code space, then it is interpreted as the occurrence of a photon loss error and it is corrected with the unitary $\ot U_1$ performing the state transfer from the error subspace of a photon loss to the logical code words. If the measurement projects into the code space, the state underwent no-jump evolution of Eq.~\eqref{eq:no_jump_state} and has been projected back to its original form by the measurement backaction $\ot P_{\rm W}$. With a probability scaling with $(\kappa \D{t})^2$ the state $\ket{E_\ua^0}$, belonging to the compliment of the logical subspace, is interpreted as a photon loss error. However, this causes no interference with the approximate recovery process as it occurs with a probability beyond the accuracy limit of the code. Recovery by measurement is reminiscent of the quantum Zeno effect: if the time evolution of a quantum state is linear (or faster) in $\D{t}$, it can be slowed to order $\D{t}^2$ by frequent measurements projecting to the non-evolved basis~\cite{OpenQuantumSystems}. The no-jump evolution scales linearly in $\kappa \D{t}$ and therefore can be corrected by frequent projective measurements. Although the recovery process $\mathcal{R}_{\rm m}$ is conceptually simpler, sophisticated measurements such as $\ot P_{\rm W}$ are presumably harder to realize at high fidelity than parity measurements~\cite{Sun14} (at least with the current technological capabilities).

\subsection{Approximate quantum error correction to $L$th order in $\kappa \D{t}$}

Here, we construct an explicit recovery process in terms of projective measurements and unitary operations, which generalizes the above discussion to multiple photon losses and shows that the binomial codes can protect against the continuous-time dissipative evolution of Eqs.~(\ref{eq:rho:el}-\ref{eq:kraus}) to an accuracy of $(\kappa \D{t})^L$. Let us take the binomial code words $\ket{W_\sigma}$ protected against $L$ photon loss errors, and choose $G=0$ and $S=L$ in Eq.~\eqref{eq:binomial}. Here we show that they are protected against the no-jump evolution to the desired accuracy as well. Our derivation is similar to Ref.~\cite{Grassl_2015}. However we arrive at a result that is not obvious from Ref.~\cite{Grassl_2015} since we exploit the explicit structure of the error operators $\ot E_\ell$~\eqref{eq:kraus}.
 
The Kraus operators $\ot E_{\ell>{L}}$ can be ignored as they have an effect of $\mathcal{O}[(\kappa \D{t})^{L+1}]$. We also ignore the parts of the remaining Kraus operators that are irrelevant to the desired accuracy and split the rest into two parts for convenience,
\begin{equation}
\ot E_\ell=\ot B_\ell+\ot C_\ell+\mathcal{O}\left[(\kappa \D{t})^{L+\frac 12}\right], \label{eq:BCsplit}\\ 
\end{equation}
for $0<\ell\le L$. We denote $\ot B_\ell$ as the important leading-order part and $\ot C_\ell$ the relevant sub-leading part,
\begin{subequations}\label{eq:EBC}
\begin{align}
   \ot B_\ell &=\ot a^\ell \sum_{\mu=\ell}^{L} \ot E_{\mu,\ell} (\kappa \D{t})^{\frac{\mu}{2}} \label{eq:EBC:b}, \\ \ot C_\ell &=\ot a^\ell \sum_{\mu=L+1}^{2L-\ell} \ot E_{\mu, \ell} (\kappa \D{t})^{\frac{\mu}{2}}, \label{eq:EBC:c}
\end{align}
\end{subequations}
where for clarity we have separated the common photon loss term. The pure no-jump evolution is handled differently~\cite{beny_perturbative_2011} and we write $\ot E_0=\exp(-\kappa \D{t} \ot n /2)=\ot B_0 + \mathcal{O}[(\kappa \D{t})^{L+1}]$. The term $\ot E_{\mu ,\ell}$ denotes the $\mu$th entry of the expansion of $\sqrt{(1-\ee^{-\kappa \D{t}})^\ell/\ell!}\exp(-\kappa \D{t} \ot n /2)$ in powers of $(\kappa \D{t})^{\frac{1}{2}}$. They are polynomials of $\ot n$ with highest degree $\le \mu/2$. 

With these preliminaries, the resulting error process is
\begin{align}
\mathcal{E}(\ot \rho)&=\sum_{\ell=0}^L \ot E_\ell \ot \rho \ot E^\dagger_\ell+\mathcal{O}\left[(\kappa \D{t})^{L+1}\right] \label{eq:errinacc} \\
&=\sum_{\ell=0}^L \left(\ot B_\ell \ot \rho \ot B^\dagger_\ell+\ot B_\ell \ot \rho \ot C^\dagger_\ell+\ot C_\ell \ot \rho \ot B^\dagger_\ell \right)+\mathcal{O}\left[(\kappa \D{t})^{L+1}\right].  \notag
\end{align}
Here we see that the part $\ot B_\ell\ot \rho \ot B^\dagger_\ell$ needs to be corrected exactly since it always has an effect larger than $\mathcal O[(\kappa\D{t})^{L+1}]$. It is not necessary to correct exactly the entire interference part $\ot B_\ell \rho C^\dagger_\ell$ since its contribution is partly beyond the accuracy limit. Hence, one can ignore the negligible $\mathcal{O}[(\kappa \D{t})^{L+1}]$ part of the interference terms and verify only that the effect of the remaining important part is independent of the logical code words. Together, if the error operators $\ot B_\ell$ and $\ot C_\ell$ for all $0\le \ell \le L$ satisfy the two following conditions, 
\begin{subequations}\label{eq:aqecc}
\begin{align}
  \braket{W_\sigma|\ot B^\dagger_\ell \ot B_\ell|W_\sigmap}&=\beta_{\ell} \delta_{\sigma \sigmap}, \label{eq:aqecc-a}\\
  \braket{W_\sigma|\ot B^\dagger_\ell \ot C_\ell |W_\sigmap}&=\nu_\ell \delta_{\sigma \sigmap}+\mathcal{O}\left[(\kappa \D{t})^{L+1}\right], \label{eq:aqecc-b}
\end{align}
\end{subequations}
the original state can be recovered to an accuracy of $(\kappa \D{t})^L$. The first condition guarantees that the errors $\ot B_k$ can be recovered exactly, see Eq.~\eqref{eq:qecc}. The second condition shows that the effect of the interference is tolerable. 

Both $\ot B^\dagger_\ell \ot B_\ell$ and $\ot B^\dagger_\ell \ot C_\ell$ up to accuracy of  $(\kappa \D{t})^{L}$ can be written as a polynomial of $\hat n $ with the highest degree of $L$. The binomial code words protected against $L$ photon losses have equal expectation value of $\ot n^\ell$, for all $\ell \le L$ and for both of the code words, which implies that the conditions~\eqref{eq:aqecc} are satisfied for them.

Now we show that the recovery process $\mathcal{R}=\{\ot R_0, \ot R_1, \ldots, \ot R_L\}$ with the Kraus operators $\ot R_k = \ot U_k \ot \Pi_{k\bmod L+1}$ results in the original state to the desired accuracy. The photon number modulo $L+1$ is measured and the measurement result $k$ has a backaction in the form of the projection $\ot \Pi_{k\bmod L+1} $. For $k\neq 0$, the error words are $\ket{B^k_\sigma}=\ot B_k\ket{W_\sigma}/\sqrt{\beta_k}$, where $\beta_k$ is defined in Eq.~\eqref{eq:aqecc-a}. Conditioned on the measurement outcome, one applies a correction unitary $\hat U_k$ performing a state transfer between the logical code words and the error words $\ket{B^k_\sigma}$, 
\begin{align}
\ot U_k = &\sum_{\sigma}\left(\ket {W_\sigma} \bra{B^k_\sigma}-  \ket{B^k_\sigma}\bra{W_\sigma} \right) +\ot U_{\rm res} \text. \label{eq:corrk}
\end{align}
Again $\ot U_k$ is completed to a unitary operator in the entire Hilbert space by $\ot U_{\rm res}$ which is an arbitrary unitary operator on the subspace complementary to the logical and error subspaces (and different for each $k$). For $k=0$, the error word needs to be orthogonalized with respect to $\ket{W_\sigma}$: $\ket{B^0_\sigma}=\big(1-\ket{W_\sigma}\bra{W_\sigma}\big)\ot B_0\ket{W_\sigma}/\sqrt{\beta_0-|\braket{\ot B_0}|^2}$, where $\braket{\ot B_0}=\braket{W_\sigma|\ot B_0|W_\sigma}$. The correction unitary is 
\begin{align}
\ot U_0 = \sum_\sigma&\Bigg[\sqrt{1-\frac{|\braket{\ot B_0}|^2}{\beta_0}}\left(\ket {W_\sigma} \bra{B^0_\sigma}-  \ket{B^0_\sigma}\bra{W_\sigma} \right)\notag \\
&+\frac{\braket{\ot B_0}}{\sqrt{\beta_0}}(\ket {W_\sigma} \bra{W_\sigma}+\ket{B^0_\sigma}\bra{B^0_\sigma})\Bigg]+\ot U_{\rm res}\text.\label{eq:corrk0}
\end{align}
These unitary operations $\ot U_k$ correct both the photon loss $\ot a^k$ and the no-jump evolution by the rest of $\ot B_k$ in Eq.~\eqref{eq:EBC:b}. As in the $L=1$ case, the error $\ot E_\ell$ shifts the photon number by $\ell$ and the initial state has a known generalized parity meaning that $\ot R_k \ot E_\ell \ot \Pi_{0\bmod L+1}=\delta_{k\ell} \ot R_\ell \ot E_\ell \ot \Pi_{0 \bmod L+1}$, for $k, \ell=\{0,1,\ldots, L\}$. Finally, we arrive at the approximate quantum error correction recovery process with accuracy $\mathcal{O}[(\kappa \D{t})^L]$: 
\begin{align}
  \mathcal{R}(\mathcal{E}(\ot \rho))\notag=&\sum_{\ell=0}^L \ot R_\ell\ot B_\ell \ot \rho \ot B^\dagger_\ell\ot R^\dagger_\ell\notag \\ &+\sum_{\ell=0}^L\ot R_\ell\left(\ot B_\ell \ot \rho \ot C^\dagger_\ell+\text{h.c.} \right) \ot R^\dagger_\ell  +\mathcal{O}\left[(\kappa \D{t})^{L+1}\right]\notag\\ =& \ot \rho \sum_{\ell=0}^L (\beta_\ell +\nu^\ast_\ell+\nu_\ell) +\mathcal{O}\left[(\kappa \D{t})^{L+1}\right]\notag \\ =&\ot \rho +\mathcal{O}\left[(\kappa \D{t})^{L+1}\right]. \label{eq:recovery_full}
\end{align}
We have used the form~\eqref{eq:errinacc} of the error process, the conditions~\eqref{eq:aqecc} and the form of the correction unitaries~\eqref{eq:corrk}-\eqref{eq:corrk0}. The last summation is the resolution of the identity, $\sum_{\ell=0}^\infty \ot E^\dagger_\ell \ot E_\ell=\ot I$, by using Eq.~\eqref{eq:BCsplit} and ignoring terms that are $\mathcal O[(\kappa \D{t})^{L+1}]$. The residual error terms in Eq.~\eqref{eq:recovery_full} depend on the binomial code parameters $S$ and $N$. We analyze the effects of this dependence on the codes' performance in Sec.~\ref{sec:perfor}. 

In summary, we have shown that the single-mode codes protected against $L$ photon loss errors are approximate quantum error correction codes protected against the continuous-time dissipative photon loss channel to an accuracy of $(\kappa \D{t})^L$. Physically, if observation of photon loss errors up to a maximum of $L$ times yields no information on population and relative phases between the logical code words, then also the observation of no-jump errors $\le L$ times yields no information and the measurement backaction does not deform the encoded quantum information of the state. This is one of our main results as it gives an explicit construction recipe for approximate error correction codes to arbitrary order in $\kappa \D{t}$ for a damped bosonic mode. 

As discussed in Sec.~\ref{sec:binomcodes}, the binomial codes may also be used to protect against more complicated error operators, such as $\ot n$. An analysis of the code parameters required to achieve such protection in the case of dissipative Lindblad time evolution under these errors is given in Appendix~\ref{sec:errors-corr-binom}. As in the case of dephasing errors in Sec.~\ref{sec:qec}, the recovery process for more general errors will be more complicated than that for photon losses alone.

\section{Binomial code performance} \label{sec:perfor}
Ignoring possible experimental infidelities of the recovery process, the performance of a binomial code is defined by the rate of uncorrectable errors. When including several error channels, that is photon loss, photon gain and dephasing errors with rates $\kappa$, $\kappa_+$ and $\gamma$,  the exact expression for the dominant uncorrectable error depends on the relative ratio of these rates. For simplicity, here we consider just a single error channel, the photon loss channel. 

\begin{figure} 
\centering 
\includegraphics[width=0.95\linewidth]{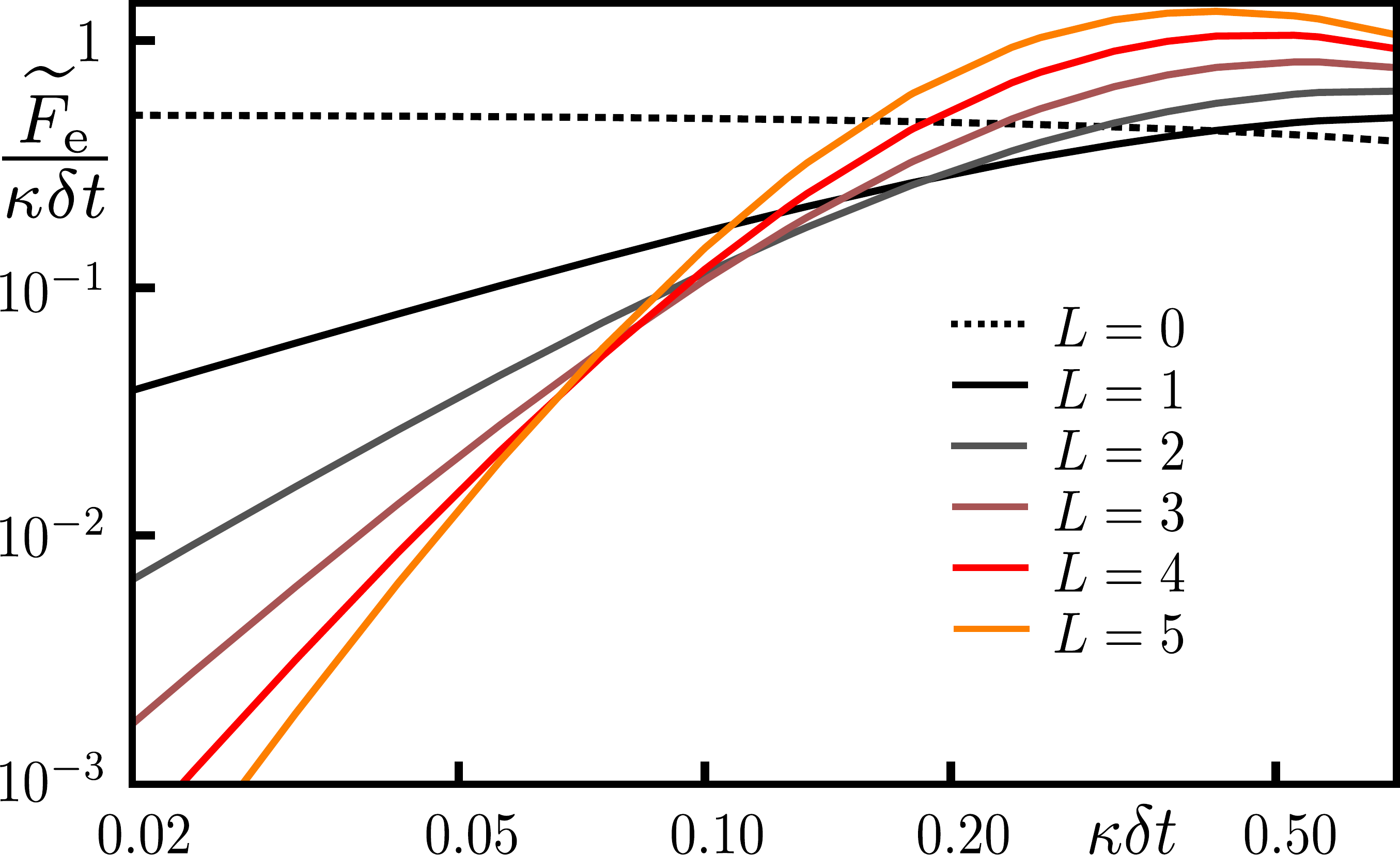}
\caption{\label{fig:scaling}The rate of entanglement infidelity $\widetilde{F}_{\rm e}/\D{t}$ for a fully mixed logical qubit state plotted as a function of the timestep $\D{t}$ (notice the logarithmic scale of both axes) for the binomial codes~\eqref{eq:binomial} with $S=N=L=1$~(black), $2$~(gray), $3$~(brown), $4$~(red) and $5$~(orange). Here we have assumed a perfectly faithful recovery process. The dashed line shows the performance of the naive encoding $L=0$, $\ket{W^{\rm i}_{\ua / \da}}=\ket{0/1}$, whose rate of entanglement infidelity at small $\D{t}$ approaches $\kappa/2$ corresponding to the rate of a photon loss with $\bar{n}=1/2$. Entanglement infidelity~\cite{Fletcher07} is calculated as $\widetilde{F}_{\rm e}=1-F_{\rm e}=1-\sum_{k=0}^L \sum_{\ell=0}^\infty|\Tr ( \ot R_k \ot E_\ell \ot \rho_{\rm c})|^2$, where $\ot\rho_{\rm c}=\frac 1 2 \sum_\sigma  \ket{W_\sigma}\bra{W_\sigma}$ is the fully mixed state of the logical code words. The entanglement infidelity for a fully mixed state is equal to the process infidelity $1-\chi_{II}$ of a quantum memory. For an ideal quantum memory $\chi_{II}=1$ and the full quantum process is just an identity operation~\cite{Chuang_and_Nielsen}. At small timestep $\D{t}$, the slopes of $\widetilde{F}_{\rm e}/\D{t}$ agree well with the slopes for the rate of the leading uncorrectable error $P_{L+1}/\D{t}$.  The binomial code with $L=1$ outperforms the naive encoding for timesteps $\D{t}\lesssim 0.4\kappa^{-1}$ and the codes with $L>1$ become favorable when $\D{t}\lesssim 0.2\kappa^{-1}$.}
\end{figure}

Let us consider the binomial code words with $S=N=L$ and study first the mean photon number $\bar{n}=\frac 1 2 \left(L+1\right)^2$. It scales quadratically with the number of protected photon loss errors $L$. This implies faster decay of the code words of higher-order protection and that to achieve the advantage of higher-order protection, the timestep $\D{t}$ must be made appropriately smaller. More precisely, the rate of uncorrectable errors is dominated by the leading uncorrectable photon loss error rate,  that is, the rate of losing $L+1$ photons during the timestep $\D{t}$: 
\begin{align}
\frac{P_{L+1}}{\D{t}}&=\frac{\braket{\ot E^\dagger_{L+1} \ot E_{L+1}}}{\D{t}}\notag\\
&\sim \kappa  (\kappa \D{t})^L\frac{\Braket{(\ot a^\dagger)^{L+1} \ot a ^{L+1}}}{(L+1)!}\sim \kappa (\kappa \D{t})^{L} L^{L+1}\,\text. \label{eq:perform}
\end{align}
This scaling result implies that for a fixed timestep $\D{t}$ there exists an optimal binomial code with finite $L$ that minimizes the uncorrectable error rate among different binomial codes. In Fig.~\ref{fig:scaling} we have demonstrated the performance of the binomial codes for $S=N=L=0,\ldots,5$ via the rate of the entanglement infidelity which, in the absence of infidelities in the recovery process and at small timesteps $\D{t}$, is well approximated by the leading uncorrectable error rate. As is clearly visible in Fig.~\ref{fig:scaling}, for a given $\D{t}$ there exists an optimal code and larger codes are preferable for smaller timesteps. 

An experimental infidelity $\eta$ related to a single recovery stage increases the error rates by $\eta/\D{t}$ favoring low-order binomial codes with longer optimal timesteps; see Appendix~\ref{app:perfor}. The optimality of a code depends also on the detailed structure of the experimental recovery process since some of the infidelities can be correctable errors suppressed by the next round of the recovery process. In addition, parity measurements often have higher fidelity~\cite{Sun14} than the unitary operations. The overall fidelity of the error recovery could be improved by making several measurements and using Bayesian estimation to increase the confidence of the error detection step~\cite{Huaixiu15}. 

If a self-Kerr non-linearity term $\propto \ot n^2$ is present in the cavity Hamiltonian, then it no longer commutes with the dissipative evolution. Such terms are difficult to avoid, either as a result of intrinsic higher-order behavior of the resonator, or due to hybridization with other non-linear degrees of freedom, such as superconducting qubits. During timesteps with only no-jump evolution, a non-linearity just introduces an additional unitary evolution that can be taken into account by defining a frame of rotating code words. In this frame the Fock state coefficients in Eq.~\eqref{eq:binomial} acquire  time-dependent complex phase factors but still obey the quantum error correction conditions~\eqref{eq:qecc}. Another possibility is to apply a gate that inverts the unitary evolution as recently demonstrated experimentally with superconducting qubit-cavity technology~\cite{Heeres15}.

During timesteps with photon jumps, the non-commutation of the Hamiltonian and $\ot a$ means precise timing of the photon jumps matters. Different Fock states acquire different, jump-time-dependent phase factors. After averaging over possible jump times, this generates additional dephasing errors, of size $\propto \D{t}$. However, the probability of a timestep with a photon jump is only  $\sim \kappa \D{t}$. As a result, the net effect is $\sim \D{t}^2$, which is higher order than the corresponding photon loss error. Thus, as long as the non-linearity is not much larger than $\kappa$, it does not break the approximate quantum error correction arguments of Sec.~\ref{sec:aqec}. Furthermore, the binomial codes can protect against additional dephasing errors by increasing the parameter $N$ in Eq.~\eqref{eq:binomial}. The cost of the higher degree of protection is an increase in the error probabilities, which can be compensated by decreasing the error correction timestep correspondingly, cf. Fig.~\ref{fig:scaling}.  The Kerr effect becomes a limiting factor for codes with large number variance $\Braket{(\ot n -\bar{n})^2}$. $\bar{n}$ only describes the `mean field' Stark shift, which does not contribute to the dephasing of Fock states relative to each other.

\section{Comparison to other codes} \label{sec:comparison}
\subsection{Two-mode codes} \label{subsec:two-mode}
As we have discussed above, even in the event of no photons being lost, the Kraus operator $\hat E_0=\exp(-\frac12\hat{n}\kappa\D{t})$ has a non-trivial effect on the single-mode code words, and so must be corrected.
This can be avoided if the words are superpositions of states with the same excitation number by combining multiple physical elements~\cite{Mabuchi96, Plenio97, Chuang97}. In particular, some of the multimode bosonic codes in Ref.~\cite{Chuang97} have the same structure as the binomial single-mode codes presented here, but are entangled across multiple photon modes,
\begin{align}
  \ket{W_{\ua/\da}}=\sum_{\substack{p \text{ even/odd}}}^{[0,N+1]}c_p\ket{p(S+1),N_{\rm tot}-p(S+1)}, \label{eq:binomial-twomodes}
\end{align}
where $N_{\rm tot}=(N+1)(S+1)$ is the excitation number, $c_p=\sqrt{\binom{N+1}{p}/2^N}$ and $\ket{n,m}$ is a state with $n$ photons in one mode and $m$ in the other. This code consists of two copies of the one-mode binomial code of Eq.~\eqref{eq:binomial} with the words entangled between the two modes.

Let us consider the simplest example, two-mode version of the single-mode code of Eq.~\eqref{eq:042}, 
\begin{align}
  \label{eq:twomode-code}
  \ket{W_\uparrow} &= \frac{\ket{0,4}+\ket{4,0}}{\sqrt2}, &  \ket{W_\downarrow} &= \ket{2,2}.
\end{align}
Assuming identical photon decay rates $\kappa$ for both modes, the Kraus evolution operator in the absence of photon losses from either mode is $\hat E_{00}=\exp(-\frac{1}{2}(\hat{n}_1+\hat n_2)\kappa\D{t})$, so that $\hat E_{00}\ket{W_\sigma} = \exp(-2\kappa\D{t})\ket{W_\sigma}$ and the code words are unchanged.
The correctable errors are still single photon losses, which can occur from either of the two modes, giving rise to different error words:
\begin{subequations}
  \label{eq:twomode-errors}
\begin{align}
  \ket{W_\uparrow} &\to \ket{E^{11}_{\uparrow}}=\ket{3,0} \text{ or } \ket{E^{21}_{\uparrow}}=\ket{0,3},\\
  \ket{W_\downarrow}& \to \ket{E^{11}_{\downarrow}}=\ket{1,2} \text{ or } \ket{E^{21}_{\downarrow}}=\ket{2,1}.
\end{align}
\end{subequations}
where $\ket{E^{i1}_{\sigma}}$ is the error word after a photon loss from mode $i$.
A parity measurement on each mode can distinguish from which mode the photon was lost, and so be used to determine whether to correct the error words $\ket{E^{11}_{\sigma}}$ or $\ket{E^{21}_{\sigma}}$.
The unitary operations required for error correction are swaps $\ket{E^{i1}_{\sigma}} \leftrightarrow\ket{W_\sigma}$, that is unitary operations 
\begin{equation}
\hat U_{i1}=\sum_\sigma \left( \ket{E^{i1}_{\sigma}}\bra{W_\sigma} - \ket{W_\sigma}\bra{E^{i1}_{\sigma}} \right)+ \ot U_{\rm res},
\end{equation}
where $\ot U_{\rm res}$ denotes an arbitrary unitary operation that completes $\hat U_{i1}$ to a unitary operation in the entire Hilbert space. These are similar to the one-mode corrections, except that they involve creating states that are entangled between the two modes. This is realizable using an experimental setup where one can generate entanglement between the modes and has sufficient separate unitary control on the individual modes. However, they are likely to have lower fidelity than the equivalent one-mode operations. See Appendix~\ref{app:hardware_twomode} for a specific hardware proposal where two cavities (or cavity modes) are dispersively coupled to a common transmon qubit with $\ot H_{\text{disp}}/\hbar=\sum_{j=1}^2\chi_j\hat{a}_j^\dagger\hat{a}_j\hat\sigma_z$, where $\hat a _j$ is the annihilation operator for the $j$th mode. If the dispersive couplings are fine-tuned to be equal $\chi_1=\chi_2$, then the codes of Eq.~\eqref{eq:binomial-twomodes} form a decoherence-free-subspace~\cite{ZanardiDFS, LidarDFS} with respect to qubit excitation induced dephasing errors $\exp(-\chi(\ot n_1+\ot n_2) \tau)$ where $\tau$ is unknown. In practice high-precision fine-tuning of $\chi_j$ is hard to achieve and one needs to correct dephasing errors with higher-order codes similar to the single-mode binomial codes. 

As in the single-mode code, the fidelity of the error correction is determined by the rate of uncorrectable errors~\cite{Leung97} and for small $\kappa\D{t}$ this is dominated by two-photon losses.
There are three paths for two-photon loss from the states of the two-mode code, Eq.~\eqref{eq:twomode-code}, compared to one path for the one-mode code, Eq.~\eqref{eq:042}. Assuming equal $\kappa$, the rate of two-photon losses via each path is the same, so the rate of uncorrectable errors for the two-mode code is three times larger than the one-mode code. Which code is preferable will depend on the fidelity of the no-jump correction for the one-mode code, as the need for this operation is eliminated in the two-mode case. More generally, for unequal $\kappa$, there will be a no-jump evolution of the form $\exp(-\frac 1 2 (\kappa_1\ot n_1+\kappa_2\ot n_2)\D{t})$ which one would have to deal with using a similar no-jump correction procedure as described for the binomial codes. 

\subsection{Cat codes}
The binomial codes are similar to existing cat codes~\cite{Cochrane99,Leghtas13,Mirrahimi14, albert_holonomic_15}. Cat codes are also approximate quantum error correction codes for a damped
bosonic mode and consist of superpositions of well-separated coherent
states, ``legs'', evenly distributed in a circle in phase space.
Cat codes with $2(L+1)$ legs protect against $L$ photon losses, and are related to the binomial codes with spacing $S=L$. In both cases, the diagnosis of errors is performed by measuring the photon
number modulo $S+1$. The four-legged cat code~\cite{Mirrahimi14}
protects against single photon losses, and so is similar to the class
of binomial codes with $L=1$, of which the simplest case is Eq.~\eqref{eq:042}.
The two logical cat code words are superpositions of coherent states $\ket{\pm \beta}$ and $\ket{\pm \ii\beta}$, 
\begin{align}
\ket{C^\beta_{\ua/\da}}&=\frac{1}{\sqrt{Z_{\ua/\da}}}\left(\ket{\beta}\pm\ket{\ii 
\beta}+\ket{-\beta}\pm\ket{-\ii\beta}\right) \notag \\
   &=\frac{1}{\sqrt{Z_{\ua/\da}}}\sum^{[0,\infty)}_{p \text{ even/odd}} 
\sqrt{\ee^{-|\beta|^2}\frac{\beta^{4p}}{2p!}}\ket{2p}\text{.} \label{eq:catcode}
\end{align}
The normalization factors $Z_{\ua/\da}$ become equal as 
$|\beta|\to\infty$. In this limit, the cat codes satisfy $\braket{C^\beta_\da|\ot n^p|C^\beta_\da} = \braket{C^\beta_\ua|\ot n^p|C^\beta_\ua}$ for all $p$, so that in the notation of Eq.~\eqref{eq:binomial} cat codes have $N\to\infty$, giving potential protection against dephasing errors to unlimited order, see Fig.~\ref{LDdfigure}. 
The difference in normalization constants for different cat states means 
that the approximate quantum error correction conditions Eq.~\eqref{eq:aqecc} are not exactly satisfied 
for generic values of $|\beta|^2$:
\begin{align}
   & \braket{C^\beta_\da|\ot E^\dagger_1 \ot E_1|C^\beta_\da} -\braket{C^\beta_\ua|\ot E^\dagger_1 \ot E_1|C^\beta_\ua}\notag\\
   &\simeq   \kappa\D{t}(\braket{C^\beta_\da|\ot n|C^\beta_\da}- \braket{C^\beta_\ua|\ot n|C^\beta_\ua})\notag \\ 
   &\simeq   4\kappa\D{t}|\beta|^2\ee^{-|\beta|^2}\left(\sin|\beta|^2+\cos|\beta|^2\right),
   \label{eq:cat-aqec-inequality}
\end{align}
where the second approximation neglects terms $\mathcal{O}(\ee^{-2|\beta|^2})$.
Similar expressions, with different trigonometric functions, can be 
found for the higher order Kraus operators.
When the right hand side of Eq.~\eqref{eq:cat-aqec-inequality} is nonzero, the cat code is 
subject to uncorrectable errors $\mathcal{O}(\kappa\D{t})$, which are suppressed 
by increasing the separation parameter $\beta$, at the cost of increasing 
the average photon number and hence the error rate.
Notice in comparison that the binomial codes with $S=1$ exactly suppress all photon loss errors to 
first order in $\kappa\D{t}$.
 
The Fock state distributions of the binomial and cat codes are binomial 
and Poissonian, respectively.
As the average number of photons is increased (larger $N$), both of these 
distributions approach a normal distribution, and so the binomial and 
cat codes asymptotically approach each other. Similarly, the qudit binomial codes approach qudit cat encoding~\cite{albert_holonomic_15} since the extended binomial coefficients also approach the normal distribution~\cite{extended_binomial}.

By construction, a photon jump event transforms one cat state into 
another cat state. To the order that the approximate quantum error conditions~\eqref{eq:aqecc} are satisfied, 
the quantum information is preserved and, as long as photon jumps are 
detected and recorded, there is no further correction needed. 
Since no-jump evolution damps coherent states, $ e^{-\kappa \delta t \hat n }| \beta \rangle = | e^{-\kappa \delta t } \beta \rangle $  $\ee^{-\kappa \D{t} \ot n }\ket{\beta} = \ket{\ee^{-\kappa \D{t}}\beta }$, the only necessary correction is a ``re-pumping'' of the cat states. 
This can be achieved using a discrete unitary correction operation (analogous to the binomial codes) or continuous non-linear amplification coming from an engineered reservoir~\cite{Leghtas13, Mirrahimi14} (analogous to similar passive/autonomous error correction schemes~\cite{Kerckhoff10,  Kapit_passive_15}). 
The basic principle  of passive schemces is stabilization of a manifold of codewords as an attractive (stable) fixed-point of drives and dissipation so that the only remaining task is to track generalized parity for photon jumps~\cite{Mirrahimi14}. 
The cat codes based on equal amplitude coherent state superpositions, \textit{cf.}~Eq.~\eqref{eq:catcode}, are `natural' candidates for these purposes since they require only gradual continuous inversion of the damping of the coherent state amplitude without active discrete correction stages. 
The two-leg cat has already been stabilized by reservoir engineering to achieve dominant two-photon drive and two-photon dissipation~\cite{Leghtas_science_15}.

The single-mode binomial codes require an explicit correction gate at every timestep whether or not a photon jump has occurred. However, binomial codes satisfy the approximate quantum error correction 
conditions to order $\D{t}$ with a smaller average photon number: 
$\bar{n}=2$ for the code of Eq.~\eqref{eq:042}, rather than 
$\bar{n}\approx 2.3$ for the cat code that minimizes 
Eq.~\eqref{eq:cat-aqec-inequality}. Furthermore, the binomial codes 
operate in a restricted Hilbert space, which could be beneficial for the 
practical construction of the unitary operators required for error 
diagnosis and recovery. This particularly applies to errors involving 
$\ot a^\dagger$ operators, whose operation on cat codes is less 
straightforward than $\ot a$ operators alone.

\subsection{Permutation-invariant codes}  
The definition of our codes, Eq.~\eqref{eq:binomial} has the same structure as the permutation-invariant codes, defined for $M$ qubits~\cite{Ouyang14_GNU, Ouyang15_GNU}:
\begin{equation}
  \label{eq:pi-code-definitions}
  \ket{\text{PI}_{\ua/\da}}=\frac{1}{\sqrt{2^{N}}} \sum_{p \text{ even/odd}}^{[0,N+1]}\sqrt{\binom{N+1}{p}}\ket{D^M_{(S+1)p}},
\end{equation}
where the Dicke state $\ket{D^M_n}$ is symmetric superposition of all permutations of $n$ up spins and $M-n$ down spins, \textit{e.g.}~$\ket{D^3_1} \propto \ket{100}+\ket{010}+\ket{001}$ in the notation of Sec.~\ref{sec:qec}.  Note that the similarity in structure is the same only for qubit code words; the qudit extension of the permutation-invariant codes~\cite{Ouyang15_GNU} is different from ours.

Although the definition of these codes in terms of excitation number take the same form, the physical and mathematical distinctions between the single mode bosonic oscillator and the many-qubit system distinguishes the goals and behavior of the two codes. The connection with the bosonic system can be identified by taking the $M\to\infty$ limit of permutation invariant codes. Then, by the Holstein-Primakoff transformation~\cite{auerbach_interacting_94}, the collective ``giant-spin'' subspace of Dicke states with total spin $M/2\to\infty$ becomes equivalent to a bosonic system. Using this mapping, we can identify several important differences between the bosonic and qubit code constructions. First, while in the spin system it is important to consider errors acting on the individual qubits~\cite{Ouyang14_GNU}, the errors in the bosonic system map to only the giant-spin operators, symmetric superpositions across all the qubit operators, \textit{e.g.}~
\begin{equation}
  \label{eq:pi-boson-mapping}
  \ot{a} \sim \lim_{M\to\infty}\frac{1}{\sqrt M}\ot{\Sigma}^- = \lim_{M\to\infty}\frac{1}{\sqrt{M}}\sum_{i=1}^{M}\ot\sigma_i^-\text,
\end{equation}
where $\ot \sigma_i^-$ is the spin lowering operator for the $i$th spin. In addition to the restriction to symmetric errors, the physical asymmetry between bosonic errors, \textit{e.g.}~$\ot a$ and $\ot a^\dagger$ makes it reasonable to consider a smaller set of errors, \textit{e.g.}~$\mathcal{\bar{E}}=\big \{\ot I, \ot a \big \}$, than the qubit case, where being able to correct any qubit error, \textit{i.e.}~$\mathcal{E}=\big \{\ot I_q, \ot \sigma^-, \ot \sigma^+, \ot \sigma_z \big \}$, is more natural. Finally, taking the $M\to\infty$ limit significantly simplifies the action of the error operators:
\begin{gather}
\begin{split}
  \label{eq:pi-clebsch-gordan}
  \frac{1}{\sqrt{M}} \ot{\Sigma}^-\ket{D^M_n} &= \sqrt{\frac{n(M-n+1)}M}\ket{D^M_{n-1}}\text{,}\\
  &\xrightarrow{M\to\infty}\sqrt{n}\ket{D^M_{n-1}}\text{.}
\end{split}
\end{gather}
The suppression of the $n^2$ term in the large $M$ limit significantly simplifies the satisfaction of the QEC conditions. As a result, the bosonic system has considerable additional flexibility in the construction of QEC codes, with concomitant performance gains. For example, for finite $M$ there is no equivalent of our smallest code, Eq.~\eqref{eq:042} in the permutation invariant codes. In general, the average excitation number in a code that corrects $L$ excitation losses scales as $L^2/2$ for bosonic binomial codes compared with  $3L^2/2$ in the permutation-invariant codes~\cite{Ouyang14_GNU}.

\subsection{GKP codes\label{subsec:GKP}}
While all codes discussed so far are defined using a discrete (\textit{i.e.},
countable) basis, the Gottesman, Kitaev, and Preskill (GKP) codes~\cite{Gottesman01} are defined using the continuous basis of non-normalizable
eigenstates of the position operator $\hat{x}$. A unique resulting
feature of GKP codes is that the correctable errors themselves form
a continuous set. The simplest ideal qubit GKP words are 
\begin{equation}
\ket{\text{GKP}_{\uparrow/\downarrow}}\propto\sum_{p\text{ even/odd}}^{(-\infty,\infty)}\hat{D}\left(p\sqrt{\frac{\pi}{2}}\right)\ket{\hat{x}=0}\,,\label{eq:gkp}
\end{equation}
where $\hat{D}$ is the displacement operator from Eq.~\eqref{eq:displacement}
and $|\hat{x}=0\rangle$ is the starting position eigenstate. In position
space, such states are infinite combs of position eigenstates spaced $2\sqrt{\pi}$ apart, so the effective spacing between the two logical states is $S=\sqrt{\pi}$.
 The same result holds in momentum
($\hat{p}$) space via Fourier transform of Eq.~\eqref{eq:gkp}. Naturally,
all position and momentum shifts $\ee^{-\ii u\hat{p}}\ee^{\ii v\hat{x}}$
with $|u|,|v|\leq\sqrt{\pi}/2$ are correctable. However, the GKP
codes can also correct any error operators expandable in the basis
of correctable shifts. Careful expansion of photon
loss $\hat{a}$ and other errors~\cite{Terhal15} has confirmed that
GKP codes can in principle (\textit{i.e.}, for small enough $\kappa \delta t$) correct photon loss, gain, and dephasing errors. 

The ideal GKP code words~\eqref{eq:gkp} contain both an infinite
number of photons and states which are perfectly squeezed in the $\hat{x}$
quadrature. To make the codes experimentally realizable, the superposition
in Eq.~\eqref{eq:displacement} has to be filtered (to keep the photon number finite)
and a distribution of states has to be substituted for the sharp $\ket{\hat{x}=0}$
state (to account for imperfect squeezing). In the traditional form
of the approximate codes, a $p$-dependent Gaussian filter is put
into the sum in Eq.~\eqref{eq:gkp} and a Gaussian wavepacket is substituted
for $\ket{\hat{x}=0}$. However, the choice of filter and starting
state can be arbitrary. There are a handful of theoretical proposals~\cite{travaglione_preparing_2002, pirandola_constructing_2004, vasconcelos_all-optical_2010} (see also Ref.~\cite{glancy_error_2006} for an error analysis of GKP states), including one based on a clever use of phase estimation~\cite{Terhal15}, to realize such approximate GKP states. GKP states are also useful in designing highly precise gates for other quantum computing architectures~\cite{Brooks13}.

The approximate code words are not perfectly orthogonal, so one must
take into account the error coming from the non-orthogonality. Since the ideal GKP states have infinite photon number, 
the approximate GKP states must contain a sufficiently high number
of photons in order to manage such imperfections. An optimistic~\cite{Terhal15} 
estimate for this photon number is $\bar{n}=4$, which uses the traditional
form of the approximate code words and bounds the error probability
coming from the non-orthogonality at about $1\%$ (Eq.~(38) of Ref.~\cite{Gottesman01}).
Using the same approximate code words, a photon number of $2$ implies
a $9\%$ error bound. As a result, the traditional form of the approximate
code words is expected to contain more photons than, \textit{e.g.}, the smallest
($\bar{n}=2$) binomial code~\eqref{eq:042}. However, the GKP code words can
protect against a larger set of errors than the minimal binomial code.
Due to the various choices of starting state and filter as well as
due to the difficulty of comparing continuous correctable error sets
to discrete ones, a detailed comparison of the relative capabilities
of the GKP and binomial/cat classes of codes remains to be done.

\section{Applications in quantum communication}\label{sec:applications}
\begin{figure}
\centering
\includegraphics[width=1.0\linewidth]{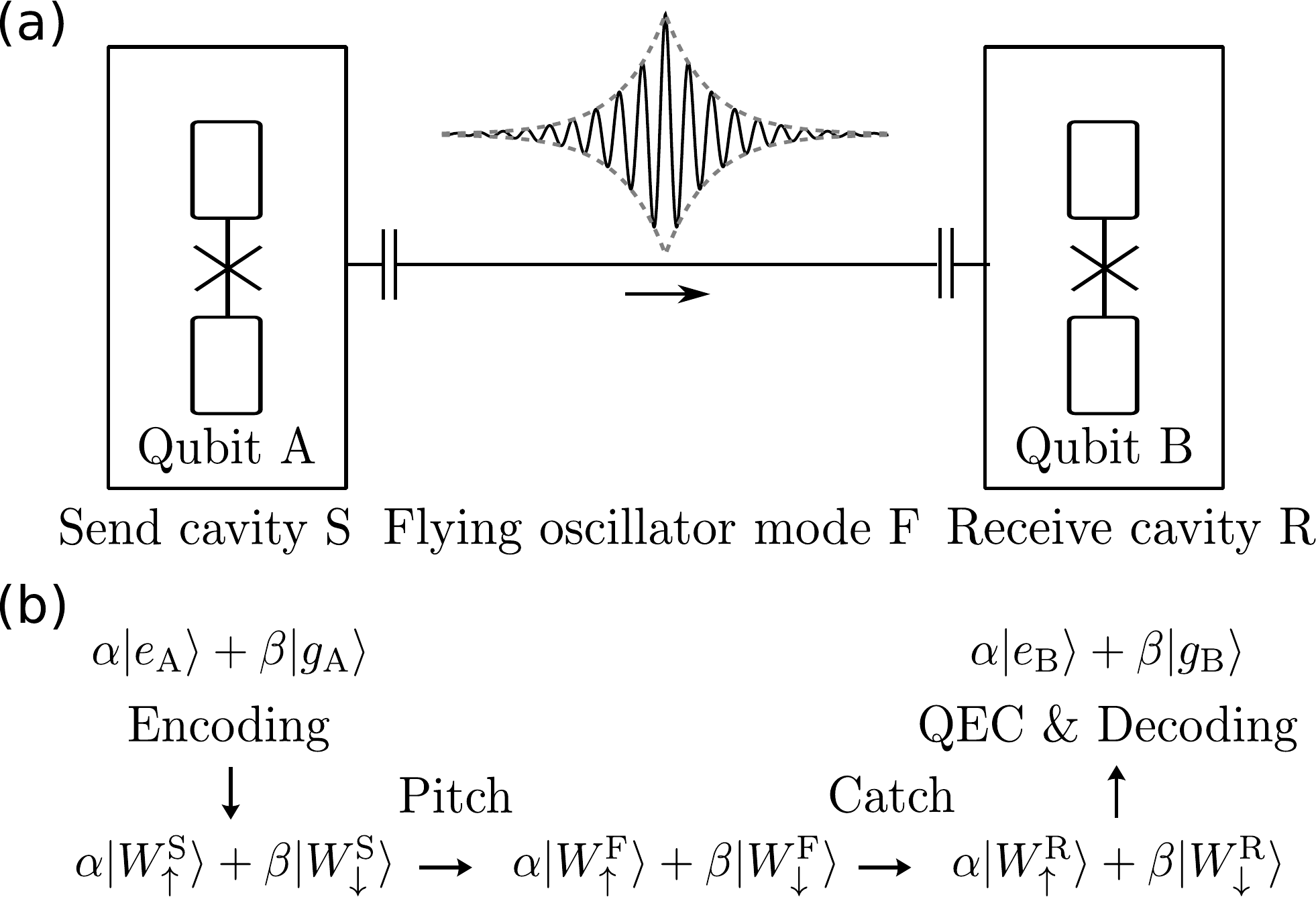} 
\caption{\label{fig:catchpitch}(a) Sketch of a circuit QED hardware proposal and (b) Schematic of a quantum state transfer scenario utilizing  quantum error correction for the  binomial quantum code words. First, the state of the qubit A is encoded in the send cavity S using code words $\ket{W_{\updownarrows}^{\rm S}}$. The state in S is allowed to leak into the flying oscillator mode F (`pitch'). By controlling the cavity decay rate, one can precisely tune a time-reversal symmetric temporal mode such that the quantum information is fully absorbed in the receiving cavity R (`catch').  Before decoding the received cavity state to the qubit B, the fidelity of the transfer process can be improved by a round of QEC in the received cavity state. }
\end{figure}

Aside from improving lifetimes of quantum memories and quantum bits, bosonic mode quantum error correction is also useful for quantum teleportation~\cite{Takeda13, Pirandola15_review} and quantum communication, which consists of quantum state transfer and generation of high-fidelity entangled pairs of quantum bits between two distant nodes in a quantum network. We consider here a primitive task, namely the `pitch-and-catch' scenario~\cite{Cirac97, Houck14, Wenner14} for quantum state transfer which can be used for quantum repeaters~\cite{Munro12, Muralidharan14}. The scenario consists of (see Fig.~\ref{fig:catchpitch}) initialization of qubit~A into a superposition of the ground and excited state, encoding this superposition into the logical code words of the send cavity via a unitary swap operation, letting the cavity state leak in a time-reversal symmetric manner (`pitch') into a transmission line or to other kind of a flying oscillator mode such that the inverse process into the receiving cavity (`catch') is most efficient~\cite{Cirac97, Korotkov11}. The transfer is finalized by decoding the received cavity state via a unitary swap operation to the qubit~B. The full process corresponds to a quantum state transfer between the qubits through the modes. The remote physical qubits can be entangled by using the same protocol with the first swap operation being replaced with a CNOT-gate between the physical qubit~A and the logical qubit of the cavity. 

The overall process is vulnerable to various errors and infidelities at the different stages of the transfer process~\cite{Vanenk97}. The most obvious imperfection is the attenuation of the state of the flying oscillator mode caused by the photon loss processes, similar to Eqs.~(\ref{eq:lind_a}-\ref{eq:kraus}), during the transmission. A crucial part of the `pitch-and-catch' process is the engineering of the temporal and spatial mode of the flying oscillator~\cite{Wenner14, Houck14, Pechal14, Pierre14, Flurin15} so that the catch by the receiving cavity is as reflection less as possible, but this is unlikely to be perfect. The pitching process can also include a conversion from microwave to optical domain or between different microwave frequencies~\cite{Andrews_Lehnert15, Konrad-Cindy14}. The fidelity of the conversion itself can be improved by using quantum error correction. In addition, there can be errors in the encoding and decoding processes between the qubits and cavities, as well the cavities and the flying oscillator mode can suffer the same loss processes we have already discussed for applications to quantum memories in Secs.~\ref{sec:qec}-\ref{sec:aqec}. If one uses the naive encoding $\ket{W^{\rm i}_{\ua/\da}}$, all these error sources lead to unfaithful quantum state transfer. When using the binomial code words or other quantum codes~\cite{Jiang09} as the logical code words in the cavities and in the transmission, the fidelity can be increased by performing a recovery process on the received cavity state before decoding it to the receiving qubit B. This way one can improve the fidelity of the process by removing the effect of the correctable errors~\eqref{eq:lll} from the full process. The performance of the different binomial codes can be estimated similarly as in Sec.~\ref{sec:perfor} and Fig.~\ref{fig:scaling} but considering \textit{e.g.}~transmission infidelity instead of the photon loss probability~$\kappa \D{t}$ during a timestep. 

\section{Controlling a logical qubit in~a~cavity} \label{sec:logical}
In addition to their use for quantum memories and communication, it would be beneficial if the encoded qubit state could be unitarily controlled to perform quantum computation. 
A naive way to realize single-qubit rotations would be to decode the cavity state back into an ancilla qubit, rotate the qubit and then re-encode the state into the cavity. 
This is not optimal since the decoherence rate of the ancilla is higher than that of the cavity \cite{ReagorMillisecondMemory,Nissim16}.

Using optimal control pulses~\cite{Khaneja05, Fouquieres11, ChenWang16, Nissim16} it should be possible to directly realize an arbitrary unitary on the cavity while minimizing the effects of the short ancilla lifetime.
For example, for the binomial code with $S=N=1$ [see Eq.~\eqref{eq:042}], the unitary corresponding to a Z gate is
\begin{align}
  \ot Z &= \ket{W_\ua}\bra{W_\ua}-\ket{W_\da}\bra{W_\da}+\ot U_{\rm res}\notag\\
  &=\ket{0}\bra{0}-\ket{2}\bra{2}+\ket{4}\bra{4}+\ot U'_{\rm res}. \label{eq:042:Z}
\end{align}On the last line we have explicitly written a particular choice of the action of $\ot U_{\rm res}$ on the non-code-word state $\ket{E^0_\uparrow}$. $\ot U'_{\rm res}$ completes the remainder of the unitary operation.
The phase gate inherits the diagonal nature of the $Z$ gate of Eq.~\eqref{eq:042:Z}:
\begin{align}
  \ot \theta &= \ee^{-\ii \theta \frac{\ot Z}{2}}\notag\\
  &=\ket{0}\bra{0}+\ee^{\ii \theta}\ket{2}\bra{2}+\ket{4}\bra{4}+\ot U_{\rm res}. \label{eq:042:phi}
\end{align}The structure of the X gate is
\begin{align}
  \ot X &=\ket{W_\da}\bra{W_\ua}+\ket{W_\ua}\bra{W_\da}+\ot U_{\rm res}\notag\\
  &=\frac{1}{\sqrt{2}}\left(\ket{2}\bra{0}+\ket{2}\bra{4} + \text{h.c.}\right)+\ot U_{\rm res}, \label{eq:042:X}
\end{align}where the functional part is an addition of $2 \bmod 4$ photons in the even photon number manifold of max $4$ photons.
The difficulty of achieving such unitaries is on the same scale as the gates needed to perform the encoding of the initial state. Consequently, the universal control of a binomial logical qubit is achieved with the same resources as the quantum error correction itself.

Joint conditional and entangling operations on two logical qubits would additionally require an entangling gate between the logical qubits. 
In Appendix~\ref{app:hardware_twomode}, we analyze a hardware proposal where this would be possible.  Progress is already underway in this direction with the recent experimental demonstration of `a cat in two boxes'~\cite{ChenWang16} which used complex entangling operations between two cavities. 

\section{Discussion}\label{sec:discuss}
So far we have considered code words constructed from Fock state superpositions with a definite generalized photon number parity, resulting in the code word spacing~$S$. This spacing readily implies a diagonal quantum error correction matrix for photon loss and gain errors. By relaxing the parity structure we can find codes with even lower rates for uncorrectable and correctable errors. However, the recovery process for these optimized codes involves more complicated measurements whose experimental fidelity is expected to be lower than that of the relatively straightforward parity measurements. We have searched for optimized codes by minimizing the largest uncorrectable error rate, the rate $P_{L+1}/\D{t}$ of losing $L+1$ photons. Using this method we have found analytic and numerical codes with reduced rates for  correctable and uncorrectable errors, some of which we were then able to transcribe analytically. As with the binomial codes, these optimized codes are exactly protected against $L$ photon losses, implying that they are approximate quantum error correction codes for the photon loss channel on accuracy~of~$(\kappa \D{t})^L$; see Appendix~\ref{app:nondiagcodes} for details.  

The recovery that is presented in Sec.~\ref{sec:aqec} for the binomial codes is not the optimal recovery; it is merely an example of a recovery process to the desired accuracy. Adjustments of the recovery process cannot beat the overall accuracy limit set by the code itself, but the prefactors of the higher-order terms in the infidelity can be made considerably smaller. A simple way of making an improvement in the sub-leading terms of the infidelity is to add to the recovery operations a unitary `echo' operation $\ot U_{\rm x}$ that performs state transfer $\ket{W_\ua} \leftrightarrow \ket{W_\da}$: $\ot R'_\ell=\ot U_{\rm x} \ot R_\ell$. The effect of this is a partial recovery of classical information from uncorrectable, higher order errors.  In general, the optimal code can be found simultaneously improving both the code and the recovery processes\----a procedure that may require numerical optimization~\cite{Fletcher07, Ng10}. 

The binomial and the optimized codes have several appealing features compared to the two-mode codes~\cite{Chuang97} and the cat codes~\cite{Cochrane99, Leghtas13, Mirrahimi14,  albert_holonomic_15} including smaller rates for correctable and uncorrectable errors, protection against photon gain and dephasing errors. It is noteworthy that our codes operate in a restricted Hilbert space  and this may be a practical advantage in designing unitary controls (in contrast to, for instance, the cat codes made out of coherent states). To achieve the full performance advantage provided by the features of the binomial codes one needs to perform sophisticated high-fidelity unitary control at a high rate. Low-fidelity unitary control favors codes with less frequent measurements and unitary operations which favors cat codes and two-mode codes. The binomial and the optimized codes outperform the cat codes and two-mode codes when the performance benefit is larger than the difference in the total infidelity of the unitary control between the codes. Current superconducting technology~\cite{Heeres15, ChenWang16, Nissim16} is on the verge of this transition.

\section{Conclusions}\label{sec:conc}
We have presented a new class of `binomial' quantum error correction codes for a bosonic mode. By constructing an explicit recovery process, we demonstrated that the binomial codes are protected to given order in the timestep against continuous dissipative evolution under loss, gain and dephasing errors. Therefore, any errors which can be expanded in terms of the creation/destruction operators of the bosonic mode can be corrected to arbitrary order. The performance of the codes is characterized by the largest rate of uncorrectable errors (\textit{e.g.}, the rate of losing $L+1$ photons for a code protecting against $L$ photon losses). Ignoring the infidelity of the recovery process, our analysis showed that with timestep $\D{t}\lesssim 0.4/\kappa$, where $\kappa$ is the photon loss rate, the naive encoding in Fock states $\ket{0}$ and $\ket{1}$ is outperformed by the smallest binomial code.  For even smaller timesteps, higher-order binomial codes become preferable. Infidelities in the recovery process favor lower-order binomial codes. 

The binomial code words consist of superpositions of equally spaced number eigenstates and are therefore eigenstates of a generalized parity. As a result, detection of loss and gain errors can be performed by measuring this generalized parity. More generally, the binomial codes, cat codes~\cite{Cochrane99,Leghtas13,Mirrahimi14, albert_holonomic_15} and Gottesman-Kitaev-Preskill (GKP) codes~\cite{Gottesman01}, share this type of structure. Namely, the logical state pairs of all three codes can be thought of as interleaved combs of eigenstates of some operator (the photon number operator for binomial/cat codes and the oscillator position operator for GKP codes) whose coefficients are related to a distribution (the binomial, Poisson, and Gaussian in the case of the binomial, cat, and GKP codes, respectively). Comparison with GKP codes suggests that it would be useful to identify commuting operators (stabilizers) for detecting errors for the binomial codes~\cite{BarbaraPrivate}. We leave this for a future work.

The generalized parity structure is a rather strong restriction on the code words and we show in Appendix~\ref{app:nondiagcodes} that, for codes built out of number operator eigenstates, better ideal performance is achieved by relaxing this structure. In future work it would be very interesting to find more examples of such codes and understand the structure of these optimized bosonic codes. Taken together, we foresee that the binomial codes and their relatives will improve the fidelity of quantum memories, communication and scalable computation based on bosonic modes. 

\begin{acknowledgments}
We are grateful for useful discussions with Huaixiu Zheng,  Reinier W.\ Heeres, Philip Reinhold, Hendrik Meier, Linshu Li, John Preskill, N.\ Read, Konrad W.\ Lehnert, Mazyar Mirrahimi, Barbara M.\ Terhal, Michel H.\ Devoret and Robert~J. Schoelkopf. We acknowledge support from the Yale Prize Postdoctoral Fellowship, ARL-CDQI, ARO W911NF-14-1-0011, W911NF-14-1-0563, NSF DMR-1301798, DGE-1122492, AFOSR MURI FA9550-14-1-0052, FA9550-14-1-0015, Alfred P.\ Sloan Foundation BR2013-049, and the~Packard Foundation 2013-39273. 
\end{acknowledgments}

\appendix

\section{Conditional unitary control of  the binomial code recovery process} \label{app:binom_unitary}
We summarize here the required conditional unitary control for the recovery of the binomial codes under the photon loss channel. As described in Sec.~\ref{sec:qec}-\ref{sec:aqec}, the binomial codes are tailored so that the photon loss errors are detected by measuring changes in the generalized photon number parity that serves as a proxy for the number of lost photons in a short timestep $\D{t}$. With superconducting circuit QED technology the ability to straightforwardly measure photon number parity stems from the strong dispersive coupling of an ancillary qubit to the cavity $\hat{H}_{\text{disp}}/\hbar=\chi\ot{\sigma}_z \ot a^\dagger \ot a$. In the strong-dispersive limit, where the strength of the dispersive coupling $\chi$ is greater than the decay rates of the qubit and the cavity, one can drive the qubit conditioned on given photon number states of the cavity~\cite{Gambetta06, Schuster07}. This can be then used for photon-number conditioned qubit operations, such as flipping the qubit state conditioned on the generalized photon parity $\ot \Pi_{ k \bmod L+1}=\sum^{[0,\infty)}_{\ell=k \bmod L+1}\ket{\ell}\bra{\ell}$:
\begin{equation}
  \hat{U}_{k \bmod L+1}= \ot \sigma_x \ot \Pi_{ k \bmod L+1} + \ot I_{\rm q}\big (\ot I-\ot \Pi_{ k \bmod L+1}\big)\text, \label{eq:cond_cav_qubit_kL}
\end{equation}
where $\ot \sigma_x$ is a Pauli matrix and $\ot I_{\rm q}$ the identity operator for the qubit. After this operation $\hat{U}_{k \bmod L+1}$, the measurement of the qubit state realizes measurement of the generalized photon parity and projection of the cavity state by $\ot \Pi_{ k \bmod L+1}$.

Error detection is followed by a correction unitary $\ot U_k$, Eqs.~(\ref{eq:corrk}-\ref{eq:corrk0}), that performs a state transfer between the error words $\ket{B_\sigma^k}$ and the logical code words $\ket{W_\sigma}$. The exact form of the correction unitary $\ot U_k$ depends on the parameters of the binomial code, Eq.~\eqref{eq:EBC}. In the strong dispersive limit, individual qubit and cavity drives are enough for implementing any unitary on the cavity~\cite{Heeres15, Krastanov15, ChenWang16, Nissim16}, see also Appendix~\ref{app:hardware_twomode}. The unitary correction applied to the cavity state can be combined with the initial conditioned unitary~\eqref{eq:cond_cav_qubit_kL} 
\begin{equation}
  \hat{U}'_{k \bmod L+1}= \ot \sigma_x \ot U_k \ot \Pi_{ k \bmod L+1} + \ot I_{\rm q}\big(\ot I-\ot \Pi_{ k \bmod L+1}\big ), 
\end{equation}
which, followed by a qubit measurement, implements the Kraus operator $\ot R_k= \ot U_k \ot \Pi_{k \bmod L+1}$ of the recovery in Eq.~\eqref{eq:recovery_full}. Repetition for all of the values of $k$ realizes the full recovery process $\mathcal{R}=\{ \ot U_k \ot \Pi_{k \bmod L+1} \}$. 

\global\long\def\a{\alpha}
\global\long\def\b{\beta}
\global\long\def\g{\gamma}
\global\long\def\c{\chi}
\global\long\def\d{\delta}
\global\long\def\o{\omega}
\global\long\def\m{\mu}
\global\long\def\n{\nu}
\global\long\def\z{\zeta}
\global\long\def\l{\lambda}
\global\long\def\e{\epsilon}
\global\long\def\x{\chi}
\global\long\def\r{\rho}
\global\long\def\t{\theta}
\global\long\def\s{\sigma}
\global\long\def\G{\Gamma}
\global\long\def\O{\Omega}
\global\long\def\L{\Lambda}
\global\long\def\P{\Phi}
\global\long\def\T{\Theta}
\global\long\def\dg{\dagger}
\global\long\def\ph{\hat{n}}
\global\long\def\aa{\hat{a}}
\global\long\def\up{\uparrow}
\global\long\def\do{\downarrow}
\global\long\def\dd{\text{d}}
\global\long\def\wt{\widetilde{W}}

\section{Moments of $\protect\ph$  for the binomial codes} \label{app:binom}
Here we show from Eq.~\eqref{eq:binomial} that the expectation value of certain moments of the photon number operator $\ph$ are identical for both code words $\ket{W_{\up/\do}}$. In other words, we show that 
\begin{equation}
\bra{W_{\sigma}}\ph^{\ell}\ket{W_{\s}}=\a_{\ell},\,\,\,\,\,\,\,\,\text{for}\,\,\,\,\,\,\,\,\,0\leq\ell\leq\max\{L,G\}\label{eq:moments}
\end{equation}
and for some real $\s$-independent $\a_{\ell}$. The $\ell=0$ case
conveniently takes care of orthonormality between the code words while
the $\ell\neq0$ conditions guarantee that the the words can be corrected
from various errors (up to the relevant order). In Appendix~\ref{app:qudit}, we extend the
definition~(\ref{eq:binomial}) to qudits and perform a similar proof
for moments of the qudit code words.

To prove Eq.~(\ref{eq:moments}), we show that the difference
of the moments of $\ket{W_{\up}}$ and $\ket{W_{\do}}$, 
\begin{equation}
\Delta_{\ell}  \equiv\bra{W_{\up}}\ph^{\ell}\ket{W_{\up}}-\bra{W_{\do}}\ph^{\ell}\ket{W_{\do}}\,,
\end{equation}
is zero. Using definition~(\ref{eq:binomial}), the difference between
the even and odd populated words is 
\begin{equation}
\Delta_{\ell}  =\frac{(S+1)^{\ell}}{2^{N}}\sum_{p=0}^{N+1}\binom{N+1}{p}p^{\ell}\left(-1\right)^{p}\,.
\end{equation}
For $\ell=0$, the sum is equivalent to a binomial expansion of $(1+x)^{N+1}$
with $x=-1$ (which is clearly zero). The nonzero $\ell$ case is
equivalent to taking derivatives of the binomial expansion and multiplying
by $x$ (before substituting $x=-1$). This is because each action
of the derivative brings down a power of $p$ while multiplication
by $x$ brings $x^{p-1}$ back to $x^{p}$. In total,
\begin{equation}
\Delta_{\ell}  =\frac{(S+1)^{\ell}}{2^{N}}\left.\left(x\frac{\dd}{\dd x}\right)^{\ell}(1+x)^{N+1}\right|_{x=-1}\,.\label{eq:delta2}
\end{equation}
Each action of the derivative acting on $(1+x)^{N+1}$ subtracts one
from the power $N+1$. Since $\ell\leq\max\{L,G\}$, the largest subtracted
power is $\max\{L,G\}$. However, since $N=\max\{L,G,2D\}$ (where
$D$ accounts for dephasing errors and is not relevant here), there
will always be a nonzero power of $1+x$ after the action
of the derivative. Therefore, the expression~(\ref{eq:delta2}) is
a polynomial in $x$ and $1+x$ containing only nonzero powers of
$1+x$. Substituting $x=-1$ into that polynomial yields $\Delta_{\ell}=0$.

An alternative basis for the binomial codes of Eq.~\eqref{eq:binomial} can be achieved by taking a normalized sum and difference of the code words $\ket{W_{\s}}$,
\begin{equation}
\ket{\widetilde{W}_{\ua/\da}}\equiv\sum_{p=0}^{N+1}\frac{(\pm 1)^{p}}{\sqrt{2^{N+1}}}\sqrt{\binom{N+1}{p}}\ket{\left(S+1\right)p}\,. \label{eq:fourbinomial}
\end{equation}
In this basis it is obvious that the moments of $\ot n$ are equal since the number distributions are identical. What is not obvious is the fact that $\braket{\widetilde{W}_\ua|\ot n^\ell |\widetilde{W}_\da}=0$. The proof is similar to the equal moments in the above case.

\section{Extended binomial qudit codes}\label{app:qudit}
We extend the above qubit states to the qudit case using extended
binomial coefficients (see~\cite{Neuschel2014, extended_binomial} and refs. therein;
these are also called polynomial coefficients~\cite{Caiado2007}).
Letting $d\geq1$ be the dimension of the logical qudit space, we
define extended binomial coefficients recursively, starting from the
ordinary binomial coefficients. Defining $\binom{n}{m}_{1}\equiv1$
and $\binom{n}{ m}_{2}\equiv \binom{n}{m}$ for non-negative integers
$n$ and $m$, the extended binomial coefficients are 
\begin{equation}
\binom{n}{m}_{d}  \equiv\sum_{k=0}^{n} \binom{n}{k}\binom{k}{m-k}_{d-1}\,.
\end{equation}
These are the coefficients of powers of $x$ in the expansion~\cite{Caiado2007}
\begin{equation}
\left(1+x+...+x^{d-1}\right)^{n}=\sum_{k=0}^{\left(d-1\right)n}\binom{n}{k}_{d}x^{k}\,.\label{eq:exp}
\end{equation}
Notice that the largest power of $x$ in such an expansion is $(d-1)n$,
which reduces to $n$ for the well-known binomial case. The last ingredient
necessary to generalize to qudits is the generalization of $\left.(1+x)^{n}\right|_{x=-1}=0$
used in the proof above. For this, we introduce the $d$th root of unity
$w\equiv\exp(\ii\frac{2\pi}{d})$ and recall that adding all powers
of $w$ from zero to $d-1$ gives zero. This reveals a set of identities
useful in defining and proving the error correction properties of
the qudit states: 
\begin{equation}
0=\left(1+w+...+w^{d-1}\right)^{n}=\sum_{k=0}^{\left(d-1\right)n} \binom{n}{k}_{d}w^{k}\,.\label{eq:orth}
\end{equation}
This sum is also zero for any nonzero power of $w$, \textit{i.e.}, $w\rightarrow w^{l}$
for nonzero integer $l$. For the zeroth power, the sum gives $d^{n}$.

We now generalize the binomial code words of Eq.~\eqref{eq:fourbinomial} to 
\begin{equation}
\ket{\widetilde{W}_{\m}}\equiv \sum_{p=0}^{\left(d-1\right)\left(N+1\right)} \frac{w^{\m p}}{\sqrt{d^{N+1}}}\sqrt{\binom{N+1}{p}_{d}}\ket{\left(S+1\right)p}\,,\label{eq:extbin}
\end{equation}
where the indices $\m,\n\in\{0,1,\ldots,d-1\}$ are from now
on evaluated modulo $d$ and $d\geq2$. Similar to the qubit case,
$S=L+G$ and $N=\max\left\{ L,G,2D\right\} $. We call these codes `extended binomial codes' as they are not to be confused with quantum polynomial codes~\cite{Aharonov1997}.

\subsection{Moments of $\protect\ph$ for the extended binomial codes}
Similar to the qubit case, it should be clear that the spacing $S+1$
between the nonzero Fock state populations of $\ket{\wt_{\m}}$ guarantees
that $\bra{\wt_{\m}}(\aa^{\dg})^\ell\aa^{\ell^{\prime}}\ket{\wt_{\n}}=0$
for all $|\ell-\ell^{\prime}|<S+1$. Therefore,
to satisfy the error correction criteria, we are once again left with
determining the powers of $\ph$ which can be used to construct any
diagonal (in Fock space) products of error operators. Here we show
that
\begin{equation}
\langle\ph^{\ell}\rangle\equiv\bra{\wt_{\m}}\ph^{\ell}\ket{\wt_{\m+\n}}=\a_{\ell}\d_{\n0}\,,\label{eq:qecapp}
\end{equation}
where $\a_{\ell}$ are real and $\m$-independent. Using definition
(\ref{eq:extbin}), we notice that 
\begin{equation}
\langle\ph^{\ell}\rangle=\frac{\left(S+1\right)^{\ell}}{d^{N+1}}\sum_{p=0}^{\left(d-1\right)\left(N+1\right)}\binom{N+1}{p}_{d}p^{\ell}w^{\n p}
\end{equation}
and the $\m$-dependence is immediately canceled. We now relate this
sum to Eq.~(\ref{eq:exp}).

For $\ell=0$, the sum is equivalent to the expansion of $(1+x+...+x^{d-1})^{N+1}$
with $x=w^{\n}$. Equation~(\ref{eq:orth}) reveals that this sum
is zero unless $\n=0$, proving that $\{\ket{\wt_{\m}}\}_{\m=0}^{d}$
are orthogonal. For the $\n=0$ case, $w^{\n}=1$ and Eq.~(\ref{eq:exp})
yields $d^{N+1}$, proving that $\{\ket{\wt_{\m}}\}_{\m=0}^{d}$ are
properly normalized. 

The nonzero $\ell$ case is equivalent to taking derivatives of the
expansion~(\ref{eq:exp}) and multiplying by $x$ (before substituting
$x=w^{\n}$). In total,
\begin{equation}
\langle\ph^{\ell}\rangle=\frac{\left(S+1\right)^{\ell}}{d^{N+1}}\left.\left(x\frac{\dd}{\dd x}\right)^{\!\ell}\,\,\,\left(1+x+\ldots+x^{d-1}\right)^{N+1}\right|_{x=w^{\n}}.\label{eq:moment}
\end{equation}
Similar to the ordinary binomial case, each action of the derivative
acting on $(1+x+\ldots+x^{d-1})^{N+1}$ subtracts one from the power
$N+1$, but $N$ is large enough so that there will always be a nonzero power of $1+x+\ldots+x^{d-1}$ remaining after the action
of all derivatives. Therefore, each term in Eq.~(\ref{eq:moment})
contains at least one nonzero power of $1+x+\ldots+x^{d-1}$. Substituting
$x=w^{\n}$ into each term yields zero unless $\n=0$ and so Eq.~(\ref{eq:qecapp})
holds.

The coefficients $\a_{\ell}$ of Eq.~\eqref{eq:qecapp} for the first few $\ell$ can be easily determined from this method~\cite{Caiado2007}:
\begin{subequations}
\begin{align}
\a_{1} & =\frac{\left(S+1\right)}{2}\left(d-1\right)\left(N+1\right),\\
\a_{2} & = \a_1 \frac{ (S+1)}{6}   
\left[\left(d-1\right)\left(3N+4\right)+2\right]\,.
\end{align}
\end{subequations}
The coefficient $\a_{1}$ is the mean photon number of the code words,
which we see scales linearly with the spacing $S$, the qudit dimension
$d$, and the maximum number of correctable errors of one type, $N$.

\section{Derivation of the Kraus operators $\ot E_\ell$} \label{app:kraus-deriv}
Here, we derive the Kraus operator representation 
\begin{equation}
  \ot \rho(t)=\sum_{\ell=0}^\infty\ot \rho_\ell (t)=\sum_{\ell=0}^\infty \ot E_\ell \ot \rho(0) \ot E^\dagger_\ell
\end{equation}
of the time evolution generated by the standard Lindblad master equation
\begin{equation}
  \dd{\ot \rho} =  \kappa\dd{t} \left ( \ot a\ot \rho \ot a^\dagger -\frac{\ot a^\dagger \ot a}{2} \ot\rho - \ot \rho \frac{\ot a^\dagger \ot a}{2} \right ). \tag{\ref*{eq:lind_a}}
\end{equation}
The zero-jump contribution consists of only the no-jump evolution under the non-Hermitian Hamiltonian $\ot{V}/\hbar=-\ii\frac{\kappa}{2}\ot a^\dagger \ot a $,
\begin{equation}
\ot \rho_0(t) = \ee^{-\frac{\kappa t}{2}\hat n} \ot \rho(0) \ee^{-\frac{\kappa t}{2}\hat n }.
\label{eq:zero-jump}
\end{equation}
The single jump contribution  $\ot \rho_1(t)$ consists of the no-jump evolution interrupted by a jump and averaged over all possible jump times, 
\begin{align}
\ot\rho_1(t)&=\int_0^t \kappa \dd{\tau} \ee^{-\frac{\kappa (t-\tau)}{2}\hat n}\ot a \ee^{-\frac{\kappa \tau}{2}\hat n }\ot \rho(0) \ee^{-\frac{\kappa \tau}{2}\hat n } \ot a^\dagger \ee^{-\frac{\kappa (t-\tau)}{2}\hat n}\notag\\
& = \left(1-\ee^{-\kappa t}\right) \ee^{-\frac{\kappa t}{2}\hat n} \ot a \ot \rho(0) \ot a^\dagger \ee^{-\frac{\kappa t}{2}\hat n},
\end{align}
where $\kappa \dd{\tau}$ is the probability for a jump during $\dd{\tau}$. We have used the identity 
\begin{equation}
 \exp\left( \kappa\D{t} \ot n\right) \ot a \exp\left(- \kappa \D{t} \ot n\right)=\ot a \exp\left(-\kappa \D{t}\right).
\end{equation}
Similarly the double jump contribution is
\begin{equation}
\ot \rho_2(t)= \frac{\left(1-\ee^{-\kappa t}\right)^2}{2!} \ee^{-\frac{\kappa t}{2}\hat n} \ot a^2  \ot \rho(0) (\ot a^\dagger)^2 \ee^{-\frac{\kappa t}{2}\hat n},
\end{equation}
and the general term for $\ell$ jumps is
\begin{equation}
\ot \rho_\ell(t)= \frac{\left(1-\ee^{-\kappa t}\right)^\ell}{\ell!} \ee^{-\frac{\kappa t}{2}\hat n}\ot a^\ell\ot \rho(0)(\ot a^\dagger)^\ell \ee^{-\frac{\kappa t}{2}\hat n},
\end{equation}
where we gather the analytic expression for the Kraus operators
\begin{align}
  \ot E_\ell=\sqrt{\gamma_\ell}\ee^{-\frac{\kappa t}{2}\ot n} \ot a^\ell =\sqrt{\gamma_\ell}\ot E_0\ot a^\ell
  =\sqrt{\gamma_\ell e^{\ell \kappa t}} \ot a^\ell \ee^{-\frac{\kappa t}{2}\ot n}. \tag{\ref*{eq:kraus}}
\end{align}
Here $\gamma_{\ell}=(1-\ee^{-\kappa  t})^\ell/\ell!$ is related to the probability of the process $\ot\rho \to \ot E_\ell \ot \rho \ot E_\ell$. When considering a small time interval $\D{t}$ and expanding $\ot E_\ell $ to the lowest order in $\kappa \D{t}$, we see that roughly speaking a photon loss error occurs with a probability amplitude proportional to $\sqrt{\kappa\D{t}}$.
 
 If this is a proper Kraus representation, it must obey the identity relation $\sum_{\ell=0}^\infty \ot E^\dagger_\ell \ot E_\ell=\ot I$. From Eq.~\eqref{eq:kraus} we have
\begin{equation}
\ot \Xi=\sum_{\ell=0}^\infty \ot E^\dagger_\ell \ot E_\ell=\sum_{\ell=0}^\infty \frac{\left(1-e^{-\kappa t}\right)^{\ell}}{\ell!} (\ot a^\dagger)^\ell \ee^{-{\kappa t}\hat n}\ot a^\ell.
\end{equation}
To see if this is the identity, we apply it to an arbitrary Fock state $\ket{m}$ and recognize that the resulting binomial expansion yields
\begin{align}
\ot \Xi\ket{m} &= \left[\sum_{\ell=0}^m  \frac{\left(1-\ee^{-\kappa t}\right)^{\ell}\left(\ee^{-\kappa t}\right)^{m-\ell}}{\ell!}\frac{m!}{(m-\ell)!} \right]\,\ket{m} \notag \\ &= \left[\sum_{\ell=0}^m \left(1-\ee^{-\kappa t}\right)^{\ell}\left(\ee^{-\kappa t}\right)^{m-\ell} \binom{m}{\ell}\right]\,\ket{m} \notag\\ &=\ket{m}. 
\end{align}
Since this is true for every $m$, the identity relation $\ot \Xi=\ot I$ is indeed satisfied. 

The Kraus operator expansion is \emph{not} unique.  This particular form organizes the errors according to how many photons are lost. Because of the no-jump evolution in between the jumps, the error operator for $\ell$ photon losses is $\ot E_\ell$ and not simply $\ot a^\ell$.  

\section{Lindblad evolution correctable by binomial codes}
\label{sec:errors-corr-binom}
Here we show how to find binomial codes that may be used to correct
non-unitary Lindbladian evolution generated by operators $\ot A_i$ of form
\begin{align}
    \tag{\ref*{eq:high-order-possible-errors}}
      \ot A_i = \sum_{jk}\xi^{(i)}_{jk}(\ot a^\dagger)^{j}\ot a^{k},
    \end{align}
occurring with respective error rates $\kappa_i$. Some of this
discussion is similar to that given for a different bosonic encoding in
Ref.~\cite{Terhal15}.
For a sufficiently small time interval $\D t$, the errors introduced by
the dissipative evolution,
\begin{align}
    \dd{\ot \rho}= \sum_i \mathcal{D}(\sqrt{\kappa_i}\ot A_i)\ot \rho \dd{t},
\end{align}
where $\mathcal{D}(\ot c)\rho=\ot c \ot \rho \ot c^\dagger - \frac 1 2 \{\ot c^\dagger \ot c,  \ot \rho \}$,  can be expanded in the parameters $\epsilon_i=\kappa_i\D t$.
For each operator $\ot A_i$, we can specify that we wish to suppress all
errors induced up to $\mathcal O(\epsilon_i^{x_i})$ where $x_i$ can be
interpreted as the maximum number of $\ot A_i$ error events in the time
interval $\D{t}$.

The values of $\epsilon_i^{x_i}$ may differ between $A_i$ and, in
general, the time evolution involves mixtures of these operators. To
represent the different combinations of $\epsilon_i$ that occur as part
of the correctable errors, we introduce the shorthand
\begin{equation}
    \label{eq:high-order-correction-order}
    \mathcal O(\gamma) = \mathcal O\left(\prod_i\epsilon_i^{y_i/2}\right)\text,
\end{equation}
where $y_i$ are any integers satisfying:
\begin{equation}
    \label{eq:high-order-yi-conditions}
    \sum_i\frac{y_i}{2x_i}\leq 1\text.
\end{equation}
To achieve this accuracy, the code words must satisfy the QEC conditions
under application of the time evolution Kraus operators $\mathcal{E} =
\{\ot E_k \}$ to the appropriate order:~\cite{Grassl_2015}
\begin{equation}
    \label{eq:high-order-qec-kraus}
    \braket{W_\sigma|\ot E_\ell^\dagger \ot E_{k}|W_\sigmap}=\alpha_{\ell
k}\delta_{\sigma \sigmap} + \mathcal O (\sqrt{\epsilon_{i}}\gamma),
\end{equation}
where $\epsilon_{i}$ may be any of the expansion parameters.

In general, it is not possible to obtain a closed form for the Kraus
operators as in Appendix~\ref{app:kraus-deriv}.
However, the evolution during a time interval $\D{t}$ can be unraveled
into different quantum trajectories~\cite{OpenQuantumSystems}. Then, for 
a given trajectory, the
system dynamics consists of continuous no-jump evolution described by
the operator:
\begin{equation}
    \label{eq:high-order-nojump}
    \ot E_0(t) = \exp\left(-\frac{t}2 \sum_i\kappa_i \ot A_i^\dagger\ot
A_i\right)
\end{equation}
and a sequence of jumps taken from the set of operators~$\sqrt{\kappa_i}\ot{A_i}$.

\subsection{Jumps alone}
\label{sec:high-order-jumps-alone}
Without the no-jump evolution, the sum over trajectories would produce
sums over all possible products of the jump operators
$\sqrt{\kappa_i}\ot A_i$, integrated over all possible jump times in the
interval $\D{t}$, to give Kraus operators
\begin{equation}
    \ot E_k \sim \prod_i\ot O_i \text{, where }{\ot O_i\in
\{\sqrt{\epsilon_i}\ot A_i\}}.\label{eq:high-order-fake-kraus}
\end{equation}
Here, we only need to consider Kraus operators that are possible up to
$\mathcal O(\sqrt{\gamma})$, as any terms that are higher order will produce
errors that are beyond the desired accuracy.
Putting these operators into the approximate QEC condition results in
conditions of the form\begin{equation}
      \label{eq:high-order-qec-bops-no-nj}
        \braket{W_\sigma|\big[  \sqrt{\epsilon_i}\ot
          A_i^\dagger\sqrt{\epsilon_j}\ot A_j^\dagger
          \dots\big]\big[ \sqrt{\epsilon_k}\ot A_k
\sqrt{\epsilon_l}\ot A_l \dots\big]|W_\sigmap}
        \propto\delta_{\sigma \sigmap}\text{,}
    \end{equation}
where the operator combinations in each square bracket can be any
possible up to $\mathcal O(\sqrt{\gamma})$. By normal ordering, these conditions
can be expressed as sums of terms like the left hand side of
Eq.~\eqref{eq:bin-conditions}.
The normal ordering procedure may produce operators that cannot be
expressed as powers of $\ot A_i$. However, following the discussion
beneath Eq.~\eqref{eq:bin-conditions}, we only need to find a binomial
code that satisfies Eq.~\eqref{eq:high-order-qec-bops-no-nj} for the terms
with the worst-case (\textit{i.e.}~largest) values of either $n_+=n_-$ 
or $|n_+-n_-|$.
All other terms, including those that may be generated by normal
ordering, then also satisfy Eq.~\eqref{eq:high-order-qec-bops-no-nj} as a
result of Eq.~\eqref{eq:bin-conditions}.

A binomial code that satisfies the worst-case instances of
Eq.~\eqref{eq:high-order-qec-bops-no-nj} can be found by choosing one
that can correct the highest order error operators:
\begin{equation}
    \label{eq:high-order-worst-error-operator}
    \ot A_i' = \left(\sqrt{\epsilon_i} \ot A_i\right)^{x_i} =
\epsilon_i^{x_i/2}\sum_{jk}\zeta^{(i)}_{jk}(\ot a^\dagger)^{j}\ot
a^{k}\text.
\end{equation}
The combinations $\ot A'^\dagger_i \ot A_j'$ contain all the worst-case
operators that appear in the conditions of
Eq.~\eqref{eq:high-order-qec-bops-no-nj}.
We then find a binomial code that can correct this error following the
rationale below Eq.~\eqref{eq:high-order-possible-errors}.

For example, consider the following (contrived) choices of $A_i$ and $x_i$:
\begin{subequations}
\label{eq:high-order-example-error-set}
\begin{align}
    \sqrt{\epsilon_1} \ot A_1 &= \sqrt{\epsilon_1} (\ot n \ot a + \ot
a^\dagger)\text, \quad &x_1 &= 1\text.
\label{eq:high-order-example-error-set-1}\\
    \sqrt{\epsilon_2} \ot A_2 &= \sqrt{\epsilon_2} \ot a\text, \quad &x_2
&= 2\text,  \label{eq:high-order-example-error-set-2}
\end{align}
\end{subequations}
In this case, from Eq.~\eqref{eq:high-order-correction-order},
$\mathcal O(\gamma) =  \mathcal O(\epsilon_1, \epsilon_2^2, \sqrt{\epsilon_1}\epsilon_2)$.
Now, $\ot A_1'=\ot A_1$ and
\begin{equation}
    \label{eq:high-order-highest-b2}
    \ot A_2' = \epsilon_2 \ot A_2^2 = \epsilon_2\ot a \ot a.
\end{equation}
To correct Eq.~\eqref{eq:high-order-highest-b2} we need $L\geq 2$,
$N\geq 2$ and Eq.~\eqref{eq:high-order-example-error-set-1} requires
$L\geq 1$, $G\geq 1$ and $N\geq 3$. These conditions give minimal
choices of $L=2$, $G=1$ and $N=3$ for a binomial code that can correct
the errors~\eqref{eq:high-order-example-error-set}.

\subsection{Including no-jump evolution}
\label{sec:including-no-jump}
Including the no-jump evolution introduces instances of the no-jump
operator, Eq.~\eqref{eq:high-order-nojump}, between the individual jump
operators to give the behavior between jump times. Since all these
intervals are $\sim \D{t}$, the no-jump operator can be expanded in terms
that are $\sim \epsilon_i$, so that the Kraus operators become
\begin{equation}
    \ot E_k \sim \prod_i\ot O_i \text{, where }{\ot O_i\in
\{\sqrt{\epsilon_i}\ot A_i,\epsilon_i\ot A_i^\dagger \ot
A_i\} }.\label{eq:high-order-fake-kraus-with-nj}
\end{equation}
Since each instance of $\ot A_i$ or $\ot A_i^\dagger$ still appears with a
factor of  $\sqrt{\epsilon_i}$, the largest combined total of $\ot a$
and $\ot a^\dagger$ operators appearing in the $\ot E_k$  to $\mathcal O(\gamma)$
is unchanged by inclusion of the no-jump evolution, and so the
appropriate $N$ can be determined as in the previous section.

However, since the no-jump evolution introduces instances of the
operator conjugates $\ot A_i^\dagger$ into the Kraus operators, it can
lead to new terms that change the number of excitations differently to
evolution under jump events alone. This will alter the required values
of $L$ and $G$.

For example, consider the choice:
\begin{equation}
    \label{eq:high-order-nj-example}
    \ot A = \ot a+\ot a\ot a,\quad x=2
\end{equation}
where the error rate $\kappa=\epsilon/\D{t}$ so that the desired order
of correction is~$\mathcal O(\epsilon^2)$.
In the presence of no-jump evolution, the set of Kraus operators includes:
\begin{subequations}
    \label{eq:high-order-bad-kraus-example}
    \begin{align}
      \ot E_0 &\approx \mathbbm{1} - \frac12\epsilon (\ot a^\dagger)^2 
\ot a + \text{(other terms)}\text,\\
      \ot E_2 &\approx\epsilon \ot A^2= \epsilon  (\ot a^4+2\ot a^3+\ot
a)\text,
\end{align}
\end{subequations}
where we have omitted some terms that are $\mathcal O(\epsilon)$ or $\mathcal O(\epsilon^2)$ because such terms will automatically satisfy the QEC conditions once we perform the analysis below. From the analysis of the previous subsection, we would consider only  $\ot A^2$, and conclude that we should use a code
with $L+G=4$.
However, putting the Kraus operators into 
Eq.~\eqref{eq:high-order-qec-kraus}, we find:
\begin{align}
\braket{W_\sigma|\ot E_0^\dagger \ot
E_{2}|W_\sigmap}=-\frac 1 2 \epsilon^2&\braket{W_\sigma|\ot a^\dagger \ot
        a^6|W_\sigmap} \notag\\&
      +\text{(other terms)}\text.
\end{align}
A binomial code satisfying Eq.~\eqref{eq:bin-conditions} for such a term
requires $L+G=5$. The no-jump evolution introduces new terms in the 
expansion of the QEC conditions that must be included.

As in the above example, the worst-case instances are introduced from
terms of the form $(\epsilon_i\ot A_i^\dagger \ot
A_i)^{\lfloor{x_i/2}\rfloor}$. We can then include them with the
worst-case error set of Eq.~\eqref{eq:high-order-worst-error-operator}:
\begin{equation}
    \label{eq:high-order-final-worst-error-operator}
    \{\ot A'_i \}= \left\{(\sqrt{\epsilon_i}\ot A_i)^{x_i}\right\} \cup
\left\{ (\epsilon \ot A_i^\dagger \ot
A_i)^{\lfloor{\frac{x_i}2}\rfloor}\right\}\text.
\end{equation}
One can now find a binomial code that corrects these errors, and as a
result the non-unitary evolution to the desired order, by expressing
these operators in the form of Eq.~\eqref{eq:high-order-possible-errors}
and satisfying the binomial code conditions for $N$, $L$ and $G$.

For the example of Eq.~\eqref{eq:high-order-nj-example}, we obtain:
\begin{equation}
\label{eq:high-order-second-example-highest-ops}
   \{\ot A'_i\} =\left\{\epsilon \left(\ot a^4+2\ot a^3+\ot a^2\right) \right\}\cup \left\{\epsilon \left(\ot a^\dagger\ot n + \ot n \ot a +\ot n^2\right)\right\}\text,
\end{equation}
which yields the parameters $L=4$, $G=1$ and $N=4$, consistent with the 
above discussion.

We note that in many physical circumstances, either $\ot A_i^\dagger\ot
A_i=f(\ot n)$, in which case the no-jump evolution does not change the
number of excitations; or the error is Hermitian, in which case
including no-jump evolution gives expressions that are the same as
Eq.~\eqref{eq:high-order-qec-bops-no-nj}. In these situations, only the error
set of Eq.~\eqref{eq:high-order-worst-error-operator} need be considered.
For example, taking the operators from 
Eqs.~\eqref{eq:high-order-example-error-set}, 
we obtain for Eq.~\eqref{eq:high-order-final-worst-error-operator}:
\begin{equation}
   \label{eq:high-order-first-example-again}
   \{\ot A'_i\} =\left\{\sqrt{\epsilon_1}\left(\ot n \ot a + \ot a^\dagger\right) , 
\epsilon_2 \ot a \ot a\right\} \cup \left\{\epsilon_2 \ot a^\dagger \ot a \right\}\text.
\end{equation}
Using this error set results in the same conditions as found without the 
no-jump evolution, Eqs.~\eqref{eq:high-order-example-error-set} and \eqref{eq:high-order-highest-b2}

\section{Performance of the binomial codes under an unfaithful recovery process} \label{app:perfor}
In Sec.~\ref{sec:perfor}, we showed that the performance of a binomial code protected against $L$ photon loss errors~\eqref{eq:binomial} is well captured by the largest rate of the unrecoverable errors, \textit{i.e.}, the rate of losing of $L+1$ photons during a timestep $\D{t}$, 
\begin{align}
\frac{P_{L+1}}{\D{t}}=\frac{\braket{\ot E^\dagger_{L+1}\ot E^\dagger_{L+1}}}{\D{t}}\sim \kappa (\kappa \D{t})^L L^{L+1}, \tag{\ref*{eq:perform}} \label{eq:performAPP}
\end{align}
where for simplicity we have assumed that $S=L$. Practically, the recovery process is always associated with an infidelity related to unfaithful gates and imprecise measurements. The total error rate $P_{\rm T}$ can be approximated by the sum of the largest unrecoverable error and the infidelity $\eta$ of the recovery process:
\begin{equation}
\frac{P_{\rm T}}{\D{t}} \simeq N\kappa (\kappa \D{t})^L L^{L+1}+\frac{\eta}{\D{t}}, 
\end{equation}
where $N$ is the prefactor of the $P_{L+1}$ scaling. The optimal timestep is $\D{t}_{\rm opt}=\kappa^{-1} \left(\eta/N L^{L+2}\right)^{\frac{1}{L+1}}$ which balances between minimizing the rate of unrecoverable errors and the infidelity of the recovery process itself. With this optimal timestep, the best performance of an unfaithfully recovered binomial code scales as function of $\eta$ as  
\begin{equation}
\frac{P^{\rm opt}_{\rm T}}{\D{t}}\simeq \kappa \eta^{\frac{L}{L+1}}(1+L)(NL)^{\frac{1}{L+1}}\sim \kappa \eta L.
\end{equation}
The performance benefit of higher order codes is achieved only with small $\eta$. 

\section{Hardware proposal for the two-mode codes} \label{app:hardware_twomode}
\begin{figure}
\centering 
\includegraphics[width=0.8\linewidth]{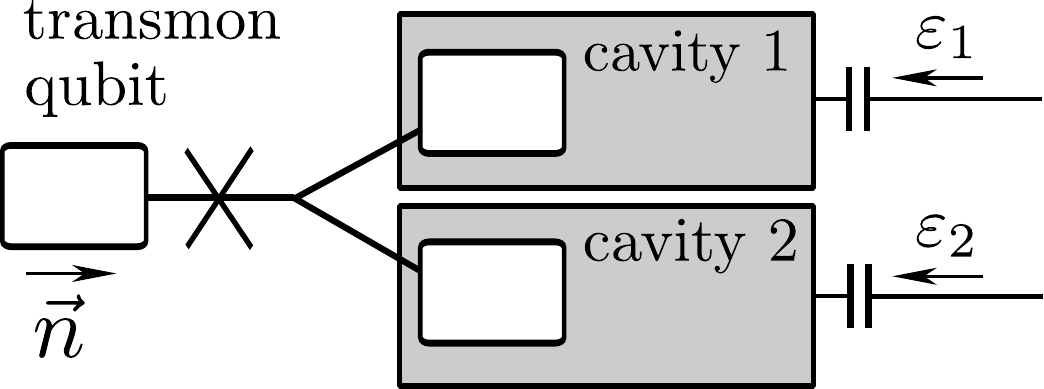}
\caption{\label{fig:twomode}Sketch of a two-cavity configuration with a dispersively coupled common transmon qubit~\cite{ChenWang16}, which is sufficient for realizing the two-mode codes. Each of the elements has a separate drive, denoted by $\varepsilon_j$ and $\vec{n}$, respectively. Alternatively, one could use two distinct modes of the same cavity.}
\end{figure}

The two-mode codes~\cite{Chuang97} or the universal control of two binomial logical qubits encoded in different individual modes could be realized using two separate modes in the same cavity or using a recently constructed system~\cite{ChenWang16} of two cavities dispersively coupled to a common transmon qubit that is used to perform unitary operations on the combined cavity system (see also a related hardware proposal in Ref.~\cite{Mirrahimi14}). Implementation requires the ability to perform the necessary measurement and error correction operations on the two cavity modes, $\hat a_j$, $j=1,2$. Here we show that the single-qubit, two-cavity experimental configuration, Fig.~\ref{fig:twomode}, is in principle sufficient to realize universal control of the two modes.

The dispersive coupling Hamiltonian is of the form $\ot H_{\text{disp}}/\hbar=\sum_{j=1}^2\chi_j\hat{a}_j^\dagger\hat{a}_j\hat\sigma_z$, where $\hat a _j$ is the annihilation operator for the $j$th mode.
Additional Hamiltonian terms come from drives on the cavities, $\ot H_{j,\text{d}}/\hbar=\varepsilon^\ast_j \hat{a}+\varepsilon_j\hat{a}^\dagger$ and the qubit $\hat{H}_Q/\hbar=\vec{n}\cdot\vec\sigma$, where the $\varepsilon_j$ and $\vec n$ are externally controlled.
The existing Hamiltonian terms can generate a more complex effective Hamiltonian using the approximate identities~\cite{Braunstein05}:
\begin{subequations}
\begin{align}
  \ee^{\ii\hat A\D t}e^{\ii\hat B\D t}\ee^{\ii\hat B\D t}\ee^{\ii\hat A\D t}&= \ee^{2 \ii(\hat A+\hat B)\D t} + \mathcal{O}(\delta t^3)\label{eq:twomode_sumidentity}\text,
  \\
    \ee^{-\ii\hat A\D t}\ee^{-\ii\hat B\D t}e^{\ii\hat A\D t}\ee^{\ii\hat B\D t}&= \ee^{[\hat A,\,\hat B]\D t^2} + \mathcal{O}(\delta t^3)\text.
  \label{eq:twomode_commidentity}
\end{align}
\end{subequations}
These identities can be applied and combined multiple times to produce superpositions of higher order commutators, \textit{e.g.}~$[\hat A,[\hat A,\hat B]]$~\cite{Braunstein05}.

To establish universal control of the multimode system, it is sufficient to show that each mode can be universally controlled, and that it is possible to generate a beamsplitter interaction $\hat{x}_j\hat{p}_k-\hat{x}_k\hat{p}_j$ (equivalent to $\hat{a}_j \hat{a}^\dagger_k+\hat{a}^\dagger_j \hat{a}_k$) between different modes $j\neq k$~\cite{Lloyd99}. 
Using the identity Eq.~\eqref{eq:twomode_commidentity}, the cavity drives along with the dispersive interaction generate effective, qubit-dependent drives on an individual cavity:
\begin{align}
   \ii\hat{H}_{j,\text{eff}}&=\bigg[\varepsilon_j\hat{a}_j-\varepsilon_j^\ast\hat{a}_j^\dagger,\,\ii\sum_{k=1}^2\chi_k\hat{a}_k^\dagger\hat{a}_k \hat \sigma_z \bigg] \notag \\
    &= \ii\chi_j\hat{\sigma}_z\left(\varepsilon_j\hat{a}_j+\varepsilon^\ast_j\hat{a}^\dagger_j\right)\text.  \label{eq:twomode_singlemode_drive}
\end{align}  
Choosing $\varepsilon_j$ to be real or imaginary results in effective operators $\propto\hat{p}_j\hat\sigma_z$ or $\hat{x}_j\hat\sigma_z$.
Combining these with pre- and post-rotations of the qubit yields, \textit{e.g.}\ $\hat{x}_j\hat\sigma_y$.
Applying Eq.~\eqref{eq:twomode_commidentity} again enables the construction of products of the mode operators~\cite{Jacobs07}, for example:
\begin{subequations}
\begin{align}
  \left[\ii\hat{x}_j\hat\sigma_y,\ii\hat{x}_k\hat\sigma_z \right] &= \ii\hat x_j \hat x_k\hat\sigma_x\text, \label{eq:twomode_xsq}\\
    \left[\ii\hat{p}_j\hat\sigma_y,\ii\hat{p}_k\hat\sigma_z \right] &= \ii\hat p_j\hat p_k \ot \sigma_x\text,  \label{eq:twomode_psq}\\
  \left[ \ii\hat{x}_j\hat\sigma_y,\ii\hat{p}_k\hat\sigma_z \right] &= \ii(\hat x_j\hat p_k+\hat p_k\hat x_j)\hat\sigma_x\text.\label{eq:twomode_xp}
\end{align}
\end{subequations}
Using Eq.~\eqref{eq:twomode_sumidentity} to sum Eqs.~\eqref{eq:twomode_xsq} and \eqref{eq:twomode_psq} with $j=k$ gives a single-mode dispersive interaction, which in combination with external cavity drives is enough to produce single-mode universal control~\cite{Krastanov15}.
Superposing Eq.~\eqref{eq:twomode_xp} with the same term with the opposite sign and $j\leftrightarrow k$ produces the beamsplitter interactions that are sufficient to give universal control of the multimode system~\cite{Lloyd99}. For practical applications, having additional, separately controlled, qubits inside each cavity may simplify the control pulses, but as this proof demonstrates, in principle they are not necessary. 

\section{Optimized bosonic codes}  \label{app:nondiagcodes}
The performance of a binomial code protected against loss of $L$ photons is dominated by the rate of losing $L+1$ photons, that is, the largest rate of uncorrectable errors. As noted in Sec.~\ref{sec:perfor}, this rate scales quite unfavorably. This is the cost of the sparse and tidy structure of the binomial codes; the occupied Fock states of the code words have definite generalized photon number parity $\{ \ket{k}\ |\ k = 0~\bmod~L+1 \}$. It is expected that by relaxing this Fock state structure we can improve the intrinsic performance of the code and pay a price in fidelity of the recovery process as the error detection and recovery will need more sophisticated unitary control. 

Here we follow the construction that we used for the binomial codes: we first find exact satisfaction of the  quantum error correction criteria for the discrete photon loss errors $\le L$ times and then demonstrate that we can find a recovery process for the continuous-time  photon loss channel to accuracy $(\kappa \D{t})^L$. In practice, we find code words $\ket{W_\sigma}$ that simultaneously minimize the dominating term of the rate for losing $L+1$ photons, 
\begin{equation} 
 \frac{P_{L+1}}{\kappa\D{t}}=\frac{\Tr\left[ \ot a^{L+1} \ot \rho_c (\ot a^\dagger)^{L+1} \right]}{(L+1)!} (\kappa \D{t})^L,
\end{equation}
where $\hat \rho_{\rm c}$ is the fully mixed state of the code words $\ket{W_\sigma}$, and satisfy the constraints of the quantum error correction criteria~\eqref{eq:qecc} for the discrete error set $\mathcal{\bar{E}}_L=\{\ot I, \ot a, \ot a^2, \ldots, \ot a^L \}$,
\begin{align}
 \braket{W_\sigma|\ot E_\ell^\dagger \ot E_{k}|W_\sigmap}=\alpha_{\ell k}\delta_{\sigma \sigmap} \text,  \tag{\ref*{eq:qecc}} \label{eq:qecc_APP}
\end{align}
for all $\ot E_{\ell,k} \in \mathcal{\bar{E}}_L$ such that $\alpha_{\ell k}$ are entries of a Hermitian matrix.  Most of the solutions we have found to this optimization problem are numerical and the detailed exploration and classification of them is left as future work. 

For the simplest error set $\mathcal{\bar{E}}_1=\{ \ot I, \ot a \}$, we have found an analytic code
\begin{subequations}\label{eq:sqrt17}
\begin{align}
  \ket{W_\ua}&=\frac{1}{\sqrt{6}}\left(\sqrt{7-\sqrt{17}}\ket{0}+\sqrt{\sqrt{17}-1}\ket{3}\right)\text,\\
  \ket{W_\da}&=\frac{1}{\sqrt{6}}\left(\sqrt{9-\sqrt{17}}\ket{1}-\sqrt{\sqrt{17}-3}\ket{4}\right)\text.
\end{align}
\end{subequations}
with remarkably low rate for the correctable error $P_1/\kappa \D{t}=\bar{n}=(\sqrt{17}-1)/2\approx 1.56$ and rate for the uncorrectable errors $P_2/\kappa \D{t}=\kappa \D{t}(3\sqrt{17}-7)/4\approx 1.34\,\kappa \D{t}$. The comparison with values $P^{\rm bin}_1/\kappa \D{t}=\bar{n}^{\rm bin}=2$ and $P^{\rm bin}_2/\kappa \D{t}=2\, \kappa \D{t}$ of the corresponding binomial code~\eqref{eq:042} shows a performance advantage both in the rate of correctable and uncorrectable errors. Due to the lack of definite parity structure with the codewords~\eqref{eq:sqrt17}, one has to use general projective measurements to detect loss of a photon instead of the straightforward parity measurements for the binomial codes (App.~\ref{app:binom_unitary}). These general projections may be realizable using current superconducting circuit technology~\cite{Heeres15, Krastanov15, ChenWang16, Nissim16} but most likely with lower fidelity than parity measurements~\cite{Sun14}.  By extending this code to be protected against a photon gain error, $\mathcal{\bar{E}}'_1=\{ \ot I, \ot a, \ot a^\dagger\}$,  we get  
\begin{subequations}\label{eq:sqrt21}
\begin{align}
  \ket{W_\ua}&=\frac{1}{\sqrt{8}}\left(\sqrt{9-\sqrt{21}}\ket{0}+\sqrt{\sqrt{21}-1}\ket{4}\right),\\
  \ket{W_\da}&=\frac{1}{\sqrt{8}}\left(\sqrt{11-\sqrt{21}}\ket{1}-\sqrt{\sqrt{21}-3}\ket{5}\right).
\end{align}
\end{subequations}
Here we observe an even more dramatic relative and absolute improvement in $P_1/\kappa \D{t}=(\sqrt{21}-1)/2\approx 1.79$ and $P_2/\kappa \D{t} =\kappa \D{t} (4\sqrt{21}-9)/4 \approx 2.33\, \kappa \D{t}$ in comparison with the values $P^{\rm bin}_1/\kappa \D{t}=3$ and $P^{\rm bin}_2/\kappa \D{t}=5.25\, \kappa \D{t}$ of the corresponding binomial code $\left( \ket{W_\ua}=(\ket{0}+\ket{6})/\sqrt{2}\text{ and }\ket{W_\da}=3\right)$.  

\subsection{Approximate quantum error correction under continuous-time dissipative evolution }
Here, we construct a recovery operation for the optimized code~\eqref{eq:sqrt17} that achieves an accuracy of $\mathcal{O}(\kappa \D{t})$ in protecting against the photon loss channel that includes both the photon loss jump and no-jump errors, as we did for the binomial code in Sec.~\ref{sec:aqec:first}. 
The optimized code~\eqref{eq:sqrt17} has a diagonal QEC matrix for the discrete errors  $\mathcal{\bar{E}}_1=\{\ot I, \ot a\}$. But for the errors~\eqref{eq:kraus} that include the no-jump evolution, $\mathcal{E}_1=\{ \ot E_0, \ot E_1 \}$, there are non diagonal elements that do not identically vanish and violate the structure of the QEC matrix due to the mixing caused by the no-jump evolution: 
\begin{align}
  \braket{W_\sigma|\ot E^\dagger_0 \ot E_1|W_\sigmap} 
&=(\kappa \D{t})^{\frac 3 2}\braket{W_\sigma| \ot n \ot a |W_\sigmap}+\mathcal{O}[(\kappa \D{t})^2]\\
&=-\delta_{\ua\da} 2(\kappa \D{t})^{\frac 32}\sqrt{5-\sqrt{17}}+ \mathcal{O}[(\kappa \D{t})^2] \notag.
\label{eq:no-jumpQECC17}
\end{align}
Notice that the diagonal approximate quantum error correction criteria~\eqref{eq:aqecc} was a result of the strict generalized parity structure of the binomial codes. Here we have deliberately broken this structure and may wonder whether the highest uncorrectable error for $\mathcal{E}_1$ is of the order of $(\kappa \D{t})^2$ or $(\kappa \D{t})^{\frac{3}{2}}$ as there is a non-vanishing term $(\kappa \D{t})^{\frac{3}{2}}\braket{W_{\ua}|\ot n \ot a|W_\da}$ in the QEC matrix.

The effect of them is most easily seen by explicitly going through the error and recovery processes. 
Remembering that a quantum state $\ket{\psi}=u\ket{W_{\uparrow}}+v\ket{W_{\downarrow}}$ transforms to $\ket{\psi_{\ell}}\equiv\hat{E}_{\ell}\ket{\psi}/\braket{\psi|\hat{E}_{\ell}^{\dagger}\hat{E}_{\ell}|\psi}^{\frac{1}{2}}$ under the action of a Kraus operator $\hat{E}_{\ell}$, the respective error states under no-jump evolution and a photon loss error are 
\begin{subequations}
\label{eq:errorpsi} 
\begin{align}
\ket{\psi_0}=&\ket{\psi}+\kappa \D{t} \left(\Gamma_0 u \ket{E_\ua^0}+\gamma_0 v \ket{E_\da^0}\right),\\
  \ket{\psi_1}=&u\left(1-\frac 12 \Gamma_0^2 |v|^2\kappa \D{t}\right) \ket{E_\ua^1} \notag \\ &+v \left(1-|u|^2 \gamma_{1} \kappa \D{t} \right)\ket{E_\da^1}+v \Gamma_0 \kappa \D{t} \ket{W_\ua},
\end{align}
\end{subequations}
to first order in $\kappa \D{t}$. The coefficients $\Gamma_{0}=\sqrt{(\sqrt{17}-3)/2}$,  $\gamma_{0}=\frac12\sqrt{3\sqrt{17}-11}$, and $\gamma_1=2/(3+\sqrt{17})$ are independent on $u$ and $v$. The normalized error words for the no-jump errors are
\begin{subequations}
\begin{align}
  \ket{E_\ua^0}& = \frac{1}{\sqrt{6}}\left(\sqrt{\sqrt{17}-1}\ket{0}-\sqrt{7-\sqrt{17}}\ket{3}\right), \\ 
\ket{E_\da^0}& = \frac{1}{\sqrt{6}}\left(\sqrt{\sqrt{17}-3}\ket{1}+\sqrt{9-\sqrt{17}}\ket{4}\right), 
\end{align}
\end{subequations}
and respectively for the photon loss errors, 
\begin{align}
  \ket{E_\ua^1}&=\ket{2}, &  \ket{E_\da^1}&=\ket{E_\ua^0},
\end{align}
where one notices that the error words overlap between the two errors $\ket{E_\da^1}=\ket{E_\ua^0}$, captured by the non-vanishing non-diagonal term $(\kappa \D{t})^{\frac{3}{2}}\braket{W_{\ua}|\ot n \ot a|W_\da}$.

We adopt now the recovery process of Eq.~\eqref{eq:recovery2}. Because we cannot use parity to detect a photon loss error we replace it with a measurement that asks whether the system is in the subspace of words after a photon loss error $\{ \ket{E_\sigma^1} \}$. This measurement performs the projection $\ot P_1=\sum_\sigma\ket{E_\sigma^1}\bra{E_\sigma^1}$. The recovery process is $\mathcal{R}=\{\ot U_0 (\ot I-\ot P_1), \ot U_1 \ot P_1 \}$, where the unitary operation $\ot U_1$ performs state transfer $\ket{W_\sigma}\leftrightarrow \ket{E_\sigma^1}$ and the unitary operation $\ot U_0$ performs the state transfer $\ket{W_\da}\leftrightarrow \ket{W_\da}+\kappa\D{t} \gamma_0\ket{E^0_\da}$ similarly as with the code~\eqref{eq:042}. Thus, we get the recovery processes
\begin{widetext} 
\begin{align}
 \mathcal{R}(\mathcal{E}(\ot \rho))=\sum_{k=0}^1 \ot R_k \left (  \sum_{\ell=0}^1 \ot E_\ell \ot \rho \ot E_\ell^\dagger \right)  \ot R_k^\dagger + \mathcal{O}\left[(\kappa \D{t})^2\right]
=&\sum_{k=0}^1 \ot R_k \left[ (1-\bar{n} \kappa \D{t})\ket{\psi_0}\bra{\psi_0} +\bar{n} \kappa \D{t} \ket{\psi_1}\bra{\psi_1}\right] \ot R_k^\dagger + \mathcal{O}\left[(\kappa \D{t})^2\right] \notag \\ 
=&\ot \rho + \ot U_1 \left[  (\kappa \D{t})^2 \Gamma_0^2 |u|^2 \ket{E_\ua^0}\bra{E_\ua^0}\right] \ot U^\dagger_1+\mathcal{O}\left[(\kappa \D{t})^2\right]\notag \\=&\ot \rho +\mathcal{O}\left[(\kappa \D{t})^2\right], 
\end{align}
\end{widetext}
where we have written the time evolution with the help of the probabilities of the Kraus operator $\Tr(\ot E_0^\dagger \ot E_0 \ot \rho)=1-\bar{n}\kappa\D{t}+\mathcal O [(\kappa \D{t})^2]$ and $\Tr(\ot E_1^\dagger \ot E_1 \ot \rho)=\bar{n}\kappa\D{t}+\mathcal O [(\kappa \D{t})^2]$ and the resulting states $\ket{\psi_i}\bra{\psi_i}$ from Eq.~\eqref{eq:errorpsi}. From this expression we see that many terms that are first order in $\ket{\psi_i}$, together with the corresponding probability, actually produce a higher order term. The second term in the second line comes from the overlap between the two errors, indicating a misidentification of the errors with a probability $\sim (\kappa \D{t})^2$. The recovery process fails to order $(\kappa \D{t})^2$ because part of the no-jump evolution is corrected as a photon loss error. However, this can be ignored as we are protecting to order $\kappa \D{t}$. Notice that the same result within the accuracy $\kappa \D{t}$ can be also achieved by the recovery $\mathcal{R}_{\rm m}=\{\ot P_{\rm W}, \ot U_1 (1-\ot P_{\rm W}) \}$. Here, the recovery error $\sim (\kappa \D{t})^2$ comes from incorrectly identifying part of the photon loss error. 

For $L=1$, the approximate QEC conditions of the continuous-time dissipative evolution under photon loss are equivalent to the QEC conditions of Eq.~\eqref{eq:qecc}. Hence, the code of Eq.~\eqref{eq:sqrt17} can correct the continuous-time error process to order $\kappa \D{t}$ as was shown above. However in general, for $L>1$, the optimized code words protecting against the continuous-time error process to order $(\kappa \D{t})^L$ will be timestep dependent. The error operators can be written as $\ot E_\ell \sim \ot a^\ell \ot E_0$. Then by writing the optimized code words $\ket{W'_\sigma}$ as $\ket{W'_{\sigma}}=\ot E_0^{-1}\ket{W_{\sigma}}/\sqrt{Z_\sigma}$ we have effectively reduced the problem to finding code words $\ket{W_\sigma}$ that satisfy QEC conditions of Eq.~\eqref{eq:qecc} for the bare photon loss errors $\mathcal{\bar{E}}_L=\{ \ot I, \ot a, \ot a^2, \ldots, \ot a^L\}$. With the  normalization factors $Z=Z_{\sigma}=\braket{W_{\sigma}|E_0^{-2}|W_{\sigma}}$, the operation $\ot E_0^{-1}/\sqrt{Z}$ is unitary for the states $\ket{W_\sigma}$ to order $(\kappa \D{t})^L$ as a consequence of QEC conditions. 


%

\end{document}